\documentclass[a4paper,11pt]{article}

\usepackage{geometry}
\usepackage{titling}
\usepackage[utf8]{inputenc}
\usepackage[T1]{fontenc}
\DeclareUnicodeCharacter{2212}{-}
\usepackage{amsmath}
\usepackage{microtype}
\usepackage{siunitx}
\usepackage{lineno}
\usepackage{authblk}
\usepackage{lscape}
\usepackage{graphicx}
\usepackage{subcaption}
\usepackage[font={footnotesize},labelfont=bf]{caption}
\usepackage[super,numbers,sort&compress]{natbib}
\usepackage{booktabs}
\usepackage{blindtext}
\usepackage{booktabs,multirow,array}
\usepackage{tikz}
\usepackage{verbatim}
\usepackage{chemfig}
\usepackage{gensymb}
\usepackage{textcomp}
\usepackage{setspace}
\usepackage{multicol}
\usepackage{pgfplots,pgfplotstable}
\pgfplotsset{compat=1.17}
\usepackage{afterpage}
\usepackage{xtab,afterpage}
\usepackage{amssymb}
\usepackage{rotating}
\usepackage{scrextend}
\usepackage[breakable]{tcolorbox}
\usepackage{xr}
\usepackage[hidelinks]{hyperref}
\hypersetup{
    colorlinks=true,
    linkcolor=black,
    filecolor=magenta,      
    urlcolor=cyan,
    citecolor=black,
    }

\usepackage{cleveref}
\usepackage{algorithm}
\usepackage[noend]{algpseudocode}
\usepackage{enumitem}
\setlist[itemize]{leftmargin=*}
\usepackage{microtype}
\usepackage{float}

\algnewcommand\algorithmicinput{\textbf{Preprocessing:}}
\algnewcommand\Preprocessing{\item[\algorithmicinput]}

\usepackage{arxiv}

\usepackage{dirtytalk}
\usepackage{txfonts}
\usepackage{soul}

\usepackage{xcolor}
\usepackage{lipsum}
\usepackage{xurl}
\usepackage{tcolorbox}
\usepackage{csvsimple}
\usepackage{tabularx}
\usepackage{longtable}

\definecolor{backcolour}{rgb}{0.95,0.95,0.92}
\definecolor{codepurple}{rgb}{0.58,0,0.82}

\begin{document}

\title{Materials science in the era of large language models: a perspective}

\makeatletter
\def\@fnsymbol#1{\ensuremath{\ifcase#1\or \dagger\or *\or \ddagger\or
   \mathsection\or \mathparagraph\or \|\or **\or \dagger\dagger
   \or \ddagger\ddagger \else\@ctrerr\fi}}
    \makeatother

\author[1]{{Ge Lei}
\thanks{These authors contributed equally.}\ \ }
\author[1, 2]{Ronan Docherty \textsuperscript{\textdagger}\ } 
\author[1]{Samuel J. Cooper \thanks{Corresponding: samuel.cooper@imperial.ac.uk}\ \ }

\affil[1]{{\textit{\footnotesize Dyson School of Design Engineering, Imperial College London, London SW7 2DB}}}
\affil[2]{{\textit{\footnotesize Department of Materials, Imperial College London, London SW7 2DB}}}

\lhead{\scshape Lei \textit{et al.}}
\chead{\scshape LLMs in materials science}
\rhead{\scshape Preprint}

\maketitle

\begin{abstract}
\begin{center}
\begin{minipage}{0.85\textwidth}
{\small Large Language Models (LLMs) have garnered considerable interest due to their impressive natural language capabilities, which in conjunction with various emergent properties make them versatile tools in workflows ranging from complex code generation to heuristic finding for combinatorial problems. In this paper we offer a perspective on their applicability to materials science research, arguing their ability to handle ambiguous requirements across a range of tasks and disciplines mean they could be a powerful tool to aid researchers. We qualitatively examine basic LLM theory, connecting it to relevant properties and techniques in the literature before providing two case studies that demonstrate their use in task automation and knowledge extraction at-scale. 
At their current stage of development, we argue LLMs should be viewed less as oracles of novel insight, and more as tireless workers that can accelerate and unify exploration across domains.
It is our hope that this paper can familiarise material science researchers with the concepts needed to leverage these tools in their own research.
}

\end{minipage}
\end{center}
\end{abstract}
\vspace{.2cm}
\begin{multicols}{2}

\section{Introduction}
\label{sec:intro}

Materials science as a discipline sits at the intersection of physics, chemistry, and often biology, and therefore requires a broad range of both skills and knowledge. A single project can cover multiple length scales, requiring various literature reviews, hypothesis generation and project planning before any experiments take place. Laboratory work can require elaborate synthesis and sample preparation routes, typically followed by a wide variety of characterization techniques. Acquired data must be processed and then analysed, either by fitting to models, comparing to simulations, or calculating uncertainties. Theoreticians must understand and leverage a variety of computational techniques from Density Functional Theory, to Computational Fluid Dynamics, and more recently to deep learning. This may require knowing multiple programming languages, as well as having the skills to deploy code across multiple environments, like high-performance clusters or cloud services. 

The rapid advancement of Artificial Intelligence (AI) - neural-network based deep-learning in particular - over the recent decade has been driven by increasingly powerful hardware and increasingly massive datasets \cite{GET_MORE_DATA}.
The culmination of this advancement is the Large Language Model (LLM), a transformer \cite{ATTN_IS_ALL_YOU_NEED} based neural network with billions of learnable parameters trained on as large a corpus of text as possible\cite{GPT_3}.
Various LLMs exist, like OpenAI's GPT-4 \cite{GPT4_REPORT}, Google's Gemini \cite{GOOGLE_GEMINI}, Meta's LLaMA 2 \cite{LLAMA}, and Anthropic's Claude 3 \cite{Claude3}. They are mostly the product of large companies with the financial and computational resources to train them, though some open source models exist \cite{FALCON, OLMO}.
Despite their simple training objective of reproducing human-like text\cite{GPT_1}, the combination of broad training data and deep network has resulted in impressive emergent capabilities and applicability to different domains and problems\cite{EMERGENT_ABILITIES_LLM}.

LLMs naturally have a strong apparent understanding of the structure of natural language, being able to translate, transpose, generate, and answer questions based on texts. They are sometimes able to (or appear able to) perform reasoning and extract patterns from textual and numerical data \cite{ICL, ICL_PROBLEMS}, extending their use beyond just language-based applications.
This combination makes them competent programmers\cite{LLMS_CODE_1}, but also effective managers or co-ordinators in complex tasks\cite{AUTOGPT}.
Whilst they perform best in workflows with a strong, LLM-independent feedback signal\cite{FUNSEARCH} they are capable of automating processes in ambiguous scenarios through trial-and-error.
Compared to say a Convolutional Neural Network (CNN), the transformer architecture is more amenable to multi-modality, able to combine and process encodings of text and images \cite{CLIP}. This multi-modality massively expands the range of problems to which LLMs can be applied \cite{GPT4_REPORT, GOOGLE_GEMINI}.

Like other computer programs, but unlike human scientists, LLMs are inexhaustible - able to run all day, every day, which is useful not just in automated digital discovery workflows, but also for setups like automated laboratories or pilot lines \cite{AUTO_CHEM_RESEARCH,RN82}. 
They are typically more flexible and adaptable than traditional computer programs, making them more effective when run continuously.
The ability to process instructions in natural language, retrieve domain knowledge, generate code and co-ordinate systems, paired with their tireless operation and immunity to boredom make LLMs appealing tools to a materials science researcher. If used judiciously they could speed up materials discovery and perform large scale analyses previously impractical for even the largest teams of researchers.


The development of the computer revolutionised information processing and research - we argue that domain-grounded LLMs will produce another step-change in materials science. 
In this paper we explore the potential role of LLMs in material science, starting with a qualitative examination of the theory underpinning transformers and LLMs in Section \ref{sec:theory}. 
Next in Section \ref{sec:capabilities} we discuss the capabilities of modern LLMs and LLM-based workflows across a variety of domains and how they might be applied to materials science. 
Section \ref{sec:case_studies} two case-studies which use LLMs in materials science workflows. The first case study uses LLMs to automate tasks during 3D microstructure analysis and the second uses LLMs to extract labels for micrographs from papers using abstracts and figure captions to create a new dataset. Finally in Section \ref{sec:issues} we examine the issues and challenges around using LLMs in research, including hallucinations, cost, and depth of understanding.

\section{LLM theory: from attention to ChatGPT}
\label{sec:theory}

\begin{figure*}
\centering
    \includegraphics[width=1\linewidth]{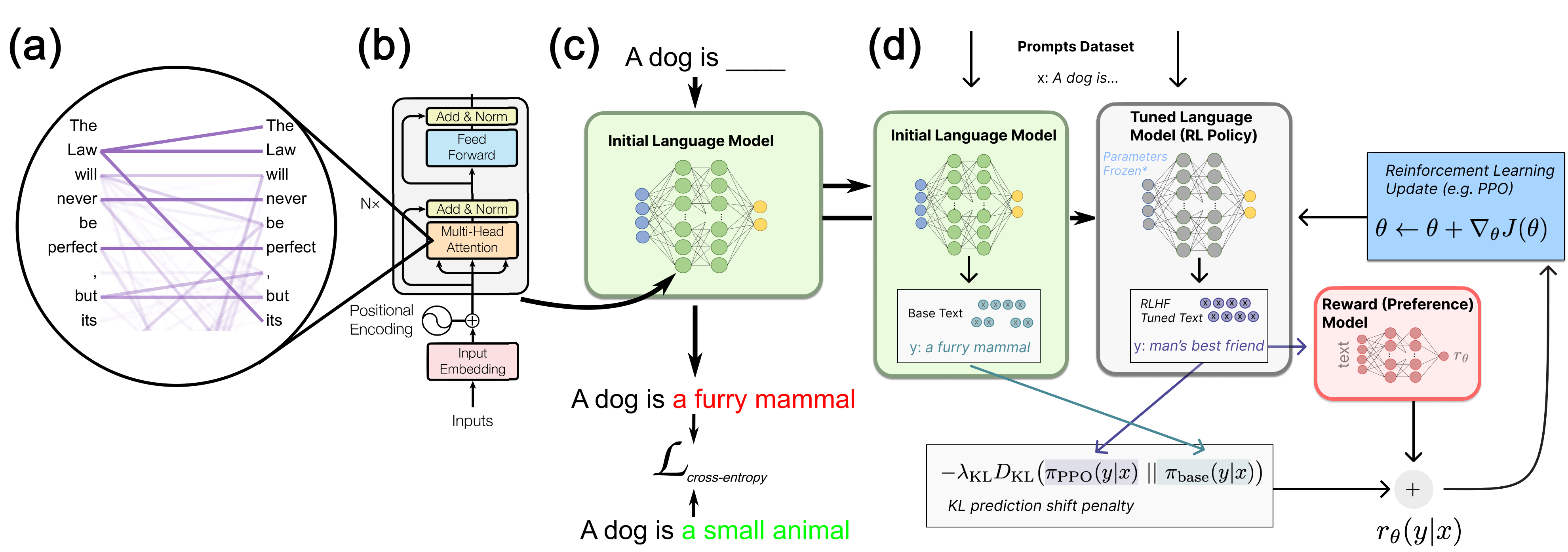}
    \caption{A multi-scale diagram of an LLM. \textbf{(a)} shows an attention map for an example sentence, note how `Law' is strongly linked to its pronoun `its'. \textbf{(b)} shows a transformer encoder layer, made up of an attention layer and (fully-connected) feed-forward layer. Multiple of these encoder layers with associated decoder layers form an LLM in \textbf{(c)}, which is pretrained in an self-supervised manner on a large text corpus. This LLM is fine-tuned to ensure its responses better match human preferences without diverging too much from the original model via RLHF, as shown in \textbf{(d)}.  Figures \textbf{(a)}, \textbf{(b)} adapted from \cite{ATTN_IS_ALL_YOU_NEED} and \textbf{(c)}, \textbf{(d)} adapted from \cite{H_FACE_RLHF}. }
    \label{fig:theory}
\end{figure*}

\subsection{Attention and transformers}
\label{sec:attention}
Attention (or self-attention), originally used for sequence modelling in recurrent neural networks\cite{RNN_ATTN}, is a mechanism designed to force a neural network to consider the rest of the elements in a sequence (the `context') in its representation of the current element. For a sentence that is represented as a sequence of tokens (efficient vector representations of words in terms of common sub-parts like prefixes) like  \say{the dog chased its own tail}, attention would place emphasis on the \say{dog} token in its representation of the \say{its} token - the consideration of context allow it to model noun-pronoun relations. An example attention map for a sentence is shown in Figure \ref{fig:theory}. A more thorough description is available in the Supplementary in Section \ref{sec:supp_attn}.

Transfomers were introduced by Vaswani \textit{et. al.} in 2017\cite{ATTN_IS_ALL_YOU_NEED} as a neural network architecture that only used self-attention for sequence modelling. The removal of recurrent layers meant less sequential operations were needed, meaning training could be parallelized even for a single training example (like a long sentence). The use of attention in place of convolutions meant shorter distances for information propagation across a sequence, making it easier to learn long-range connections. 

Despite being the most efficient way to include the whole context of a sequence of $n$ tokens in a single layer, computing the interaction of every token with every other token means attention is an $O(n^2)$ operation. This limits the total `context length' of the input sequence based on the amount of (GPU) memory. The quadratic scaling is the major downside of transformers and researchers are looking to mitigate this with techniques like windowed attention\cite{WINDOWED_ATTN} or moving to linear state-space models like Mamba\cite{MAMBA}, though these approaches lose global context. 

Another consequence of attention is that there is no implicit ordering of tokens in the network - this information must be added in the form of a `positional embedding` to the vector representation of each token in the sequence. The simplest way of doing this is word-order, \textit{i.e,} which number the token is in the sequence, though other embeddings like sinusoidal or learned embeddings are also used\cite{POSITIONAL_EMBEDDINGS}.
An embedding is just a vector representation of a quantity in a new subspace - this can be as simple as one-hot encoding showing the presence of a feature or as complicated as a set of features learned by a deep CNN.

\subsection{Pretraining and language modelling}
\label{sec:pretraining}
Supervised training is updating the weights of a neural network to minimize the loss between the labels predicted by a model, $\hat{y}$ and the labels from the dataset $y$ for a given input $x$. As an example, the $x$ could be a photo of a dog and $y$ could be a label from a human saying \say{dog}. In  training, the human labels $y$ are replaced with some transformation of the input $y = f(x)$.

Ideally during self-supervised training the network learns strong representations of the data and can be fine-tuned or paired with another network on labelled data for specific tasks. This has two advantages - firstly that it reduces the amount of human labour needed to label the inputs, $x$, and secondly that it is believed to produce more robust representations\cite{DINO} than supervised learning, due to the lack of `shortcuts' available. 
An example of a `shortcut' is learning to predict a dog by detecting a lead, or detecting a polar bear based on ice in the background - learning these might mean ignoring more relevant and generalisable features\cite{SHORTCUTS}. 

Transformers are parallelizable so scale well with added data and compute, and can easily learn long-range connections \cite{ATTN_IS_ALL_YOU_NEED}.
Self-supervised learning requires little or no human input - massive text datasets can be collected through automated web-scraping\cite{COMMONCRAWL} - and generates strong learned representations.
This combination makes transformers prime candidates for self-supervised learning on large text datasets to create multi-purpose language models. 

One of the first works to apply self-supervised learning to large text datasets with transformers was Radford \textit{et. al.} in 2018 \cite{GPT_1}, where a transformer was pre-trained on 7,000 unpublished books before being fine-tuned on tasks like question-answering and classification. It was pre-trained using next-token prediction and operated autoregressively, \textit{i.e} it predicted next-token probabilities for all tokens in its vocabulary, selected the highest one, added it to the input and predicted the new next token. This was called \say{generative pre-training}, and the model was called a \say{Generative Pretrained Transformer} (GPT). 

GPT's pre-training was left-to-right causal language modelling where the sequence had to be masked to prevent the transformer seeing future tokens (specifically the current token of interest) and predicting that. An alternative approach is masked language modelling, where only the current token of interest in the sequence is masked and the rest is part of the context - this is bidirectional and future context can be considered. This was the approach used for Google's BERT in 2018 \cite{BERT}. The bidirectional language modelling meant BERT had higher performance on benchmarks but meant it could not be autoregressive/generative - a key factor in ChatGPT's later popularity.

\subsection{Aligning outputs via RLHF}
\label{sec:RLHF}
Always selecting the most probable token during autoregression leads to coherent and deterministic results, but can limit the ability of the model to be `creative'. Picking tokens in proportion to their probability is a simple strategy for more diverse text, and a common way to parameterise the distribution and therefore control text generation is `temperature'\cite{TEMPERATURE}. Temperature is a scalar term introduced before the softmax function that generates per-token probabilities, with a large temperature increasing the probability of previously rare tokens and making more common tokens less likely\cite{TEMPERATURE_2, TEMPERATURE_3}. A low temperature does the opposite, and a temperature of 0 is used to refer to most likely token selection.

`Prompting' is a consequence of the autoregressive learning objective of LLMs - a user's prompt is given to the LLM as a sequence and the LLM generates the most likely subsequent tokens. The model must be fine-tuned to act in a true question/answer or chatbot style\cite{GPT_3}. The notion of prompting has found success in other domains, like Meta's promptable `Segment Anything Model'\cite{SAM}.

In 2022 OpenAI published a paper on \say{InstructGPT}\cite{RLHF_GPT}, a pre-trained model which was then trained on a dataset of prompts to desired responses and finally fine-tuned via RLHF\cite{RLHF}.
RLHF, shown in Figure \ref{fig:theory}, contains two LLMs - a frozen LLM and the LLM to fine-tune. A prompt is fed to both, and the fine-tuned LLM's response is fed to a reward model (a NN trained to emulate human preferences) to generate a reward score. A second term is added to the reward based on the KL divergence between the frozen and fine-tuned LLM to prevent model drift. This reward is fed into a reinforcement learning policy, like Proximal Policy Optimization (PPO)\cite{PPO} to update the weights.

InstructGPT had significantly fewer parameters than GPT-3 but outperformed it, signalling the power of reinforcement learning in aligning a model's outputs with human preferences. Despite this impressive performance it is worth noting that at no point during the pre-training, training or fine-tuning are models explicitly trained to minimize factual errors or to reason - saying the sky is green goes against human preference and would therefore be penalised, but most labellers would be unaware if the model had confused ferro- and ferrimagnets if the text was otherwise coherent.

\section{Capabilities of LLMs in research}
\label{sec:capabilities} 
Machine learning has seen widespread application in materials science, from characterization\cite{ML_CHARACTERIZATION, CONCRETE_ML_CHARACTERIZATION} to property prediction\cite{ML_PROPERTY_PRED_1, ML_PROPERTY_PRED_2, ML_PROPERTY_PRED_3} to materials discovery and design\cite{ML_MOLECULE_DESIGN, SCALE_DL_MATERIALS_DISCOVERY, MATTERGEN}.
They have mostly been applied in well-structured tasks with strong supervision (\textit{i.e,} fully labelled single-task datasets)\cite{HACKATHON_LLM}.
Frequently this has involved researchers developing models trained only on their data for their problems, causing poor generalisation to new materials or processing conditions\cite{MATERIALS_DATA_DIVERSITY}.

LLMs, by virtue of the the size of the number of parameters and the scale of their training data, have strong natural language skills and emergent properties that make them promising candidates for processing more unstructured and varied data\cite{HACKATHON_LLM}.
In this section we explore some of these emergent capabilities, examine how researchers have used them in various disciplines and consider them in a materials science context.

\subsection{LLM properties: intrinsic and emergent}
\label{sec:properties}

\begin{figure*}
\centering
    \includegraphics[width=0.75\linewidth]{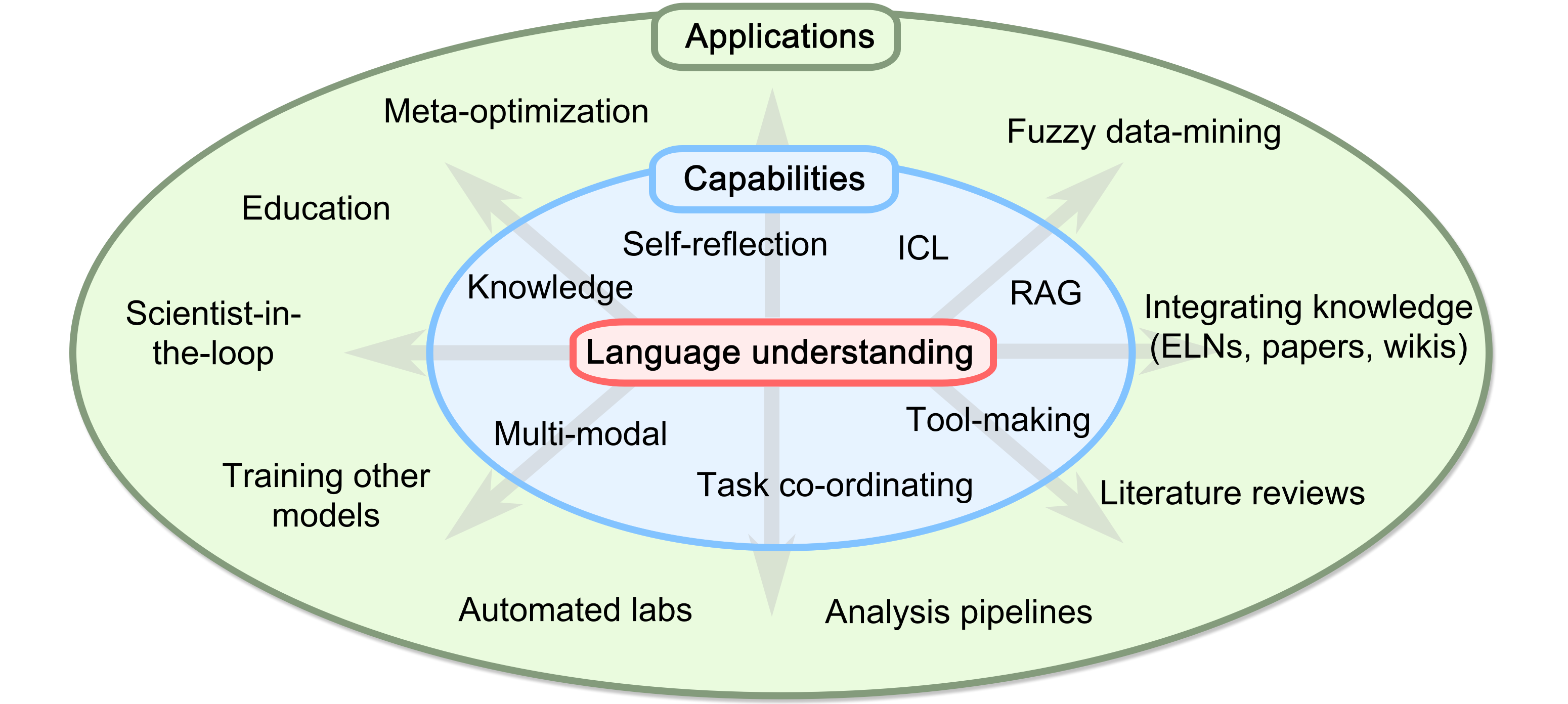}
    \caption{Diagram of LLM capabilities explored in Section \ref{sec:capabilities} and potential materials-science related applications. These emergent capabilities can be combined with each other and integrated into traditional pipelines (genetic algorithms, databases, \textit{etc.}) to form the different applications.}
    \label{fig:properties}
\end{figure*}

\subsubsection*{Optimizing responses with prompt engineering}
\label{sec:prompt_enginering}
Prompt engineering refers to optimizing the prompt the user gives to an LLM in order to produce a `better' answer or response. This can involve making the response more or less precise, conform to specific rules or schema changing the level of the explanation, \textit{i.e,} \say{explain like I'm five years old...}. 
Choosing the correct prompt is more of an art than a science, but some work has been done on creating and testing various principles\cite{PROMPT_CATALOG, PROMPT_PRINCIPLES} or using an LLM to optimise prompts for another LLM\cite{LLM_PROMPT_LLM}. It has been found that it is generally better to be precise, structured (using paragraphs, whitespace and question/answer blocks) and explain to the LLM what `role' it should act in. 
Example prompts for our case studies, which utilise these principles, are in Supplementary Sections \ref{sec:supp_case_1} and \ref{sec:supp_case_study_2_prompts}.

\subsubsection*{In-Context Learning (ICL) and property prediction}
One important emergent property that is useful to optimize a prompt is in-context learning, or few-shot learning\cite{GPT_3}, where a few example input/output pairs are provided to the model in the prompt. This is useful for familiarising the model with unknown concepts. 
Why it works is still a topic of debate, and it is a counter-intuitive phenomenon. Min \textit{et. al.}\cite{ICL_IS_WEIRD} found that during in-context learning, randomly swapping output labels in examples the (introducing wrong information into the context) only slightly decreased the accuracy of LLMs when predicting related new samples. This suggested showing the model the structure and input/output distributions were more important than demonstrating the underlying logical mapping. 

However, subsequent research\cite{ICL} studied this effect with larger LLMs and found that as the number of model parameters increased, the adverse effects of wrong information in examples became more pronounced.  As the percentage of swapped labels increased, the accuracy on unswapped examples dropped below $50\%$ (the random baseline), implying the larger LLMs learnt the reverse logical mapping from the (swapped) in-context examples, rather than just the problem structure. 

Google's Gemini 1.5 Pro\cite{GOOGLE_GEMINI_1.5} recently demonstrated its advanced in-context learning capabilities by accurately translating English into Kalamang, a language spoken by fewer than 200 people. Initially, the model struggled with translations due to a lack of Kalamang training data. However, after processing 250k tokens of linguistic documentation on Kalamang without undergoing a traditional training regimen, it achieved near-human levels of translation accuracy.


As well as various natural language and mathematical problems\cite{ICL_PROBLEMS}, ICL has been used for both quantitative and qualitative material property prediction \cite{GPT_MOLBERTA, LLM_PROP, BAND_GAP_PRED_LLM, LLM_SCIENTIFIC_DISCOVERY, HACKATHON_LLM}. However, Microsoft AI4Research noted that despite good qualitative predictions, the quantitative predictions of LLMs were lacking\cite{LLM_SCIENTIFIC_DISCOVERY}. 

\subsubsection*{Error correction via Chain-of-Thought (CoT) reasoning \& self-reflection}
\label{sec:chain_of_thought}
Chain-of-Thought reasoning involves asking the LLM to break a problem into a series of steps in order to improve accuracy. This can be included in the prompt in a phrase like \say{think through the problem step-by-step} or as part of an ICL prompt where the example given is broken into a series of steps. CoT has been found to improve performance on tasks that require complex reasoning\cite{COT_REASONING, REACT_COT}. 

Various explanations for this improved performance have been suggested, including that requesting the model think step-by-step increases the length of the sequence. Recall that during autoregression the whole sequence including the current output is fed into the model to generate the next token - adding more tokens gives the model more context and thus more `space' to compute with, as more text means more interactions in the attention layer. 
Another possibility is that longer sequences reduces the space of likely sequences to those that contain the correct answer; if the model has repeated \say{John has 4 apples} multiple times as part of its explanation the probability of outputting future tokens that use (directly or indirectly) a different number of apples is reduced compared to directly outputting the answer. 

Self-reflection is a consequence of ICL and CoT and involves giving an LLM an evaluation of its previous prompt in a new prompt, this can be pointing out errors or a broader evaluation. This can be from a human\cite{COT_REASONING}, a program (\textit{i.e,} a stack trace)\cite{COT_VERBAL_RL}, another LLM\cite{COT_VIA_LLMS} or even itself\cite{COT_LLM_SELF_IMPROVE}. Self-reflection improves performance, potentially for the same reasons as ICL and CoT, but also because correcting a wrong output may be a simpler task than generating the correct output \textit{de novo}. 

\subsubsection*{Pre-existing and fine-tuned materials domain knowledge}
\label{sec:knowledge}
LLMs are trained on large corpuses of text that contain facts about the world, including large datasets of scientific papers\cite{THE_PILE}. Common facts will be repeated many times across these texts, making it statistically likely that an LLM will reproduce them when prompted to. Microsoft AI4Science found that \say{In biology and materials design, GPT-4 possesses extensive domain knowledge}, which they evaluated by asking domain experts to rate outputs about various drug molecules, general materials design principles, mathematical concepts like PDEs and more\cite{LLM_SCIENTIFIC_DISCOVERY}. 

The ability to act as an oracle for common shallow information across many domains\cite{LLMS_KNOWLEDGE_BASE} is useful in a multi-disciplinary field like materials science, but the regression to the mean encouraged by the pre-training task and autoregression can limit the usefulness of LLMs for deep information recall. 

One way of overcoming this is fine-tuning on domain-specific knowledge. This domain specific knowledge can be collected traditional web-scraping or using ML models\cite{NLP_MAT_SCI_ANALYSIS} and then used to fine-tune a language model like BERT\cite{BERT}. Models like SciBERT\cite{SCIBERT} outperformed BERT and other state-of-the-art models for tasks like text classification or Named Entity Recognition (NER). MatSciBERT\cite{MATSCIBERT} took this process a step further and fine-tuned SciBERT on materials science specific data to outperform SciBERT on materials science text tasks.

Full fine-tuning of any large ML model is expensive and risks `catastrophic forgetting'\cite{CATASTROPHIC_FORGETTING}, where a model loses information from its general (pre)training during the domain-specific fine-tuning. One way to alleviate both the cost and catastrophic forgetting problem is Parameter-Efficient-Fine-Tuning (PEFT), where only a small subset of the model's parameters are updated. Examples PEFT schemes include LORA\cite{LORA}, adapters\cite{LLM_ADAPTERS} and prompt/prefix-tuning methods\cite{PROMPT_TUNING, PREFIX_TUNING}.


\subsubsection*{Comprehensive programming skills}
\label{sec:programming}
Large quantities of text exist online (and therefore in LLM training sets) about programming: discussions, help forums and source code, and the move towards approachable, high-level programming languages like Python means source code is increasingly similar to natural language. 
These two facts mean LLMs are proficient at generating, modifying, correcting and summarizing code in a variety of languages for a variety of tasks\cite{LLMS_CODE_1, LLMS_CODE_2}.

Programming is ubiquitous in modern science, from data processing, analysis, visualization, simulations, instrument interfaces, \textit{etc.} and the ability to write reasonable code across all these different tasks is obviously useful for researchers. LLMs have been shown to be proficient in these tasks in a materials science context\cite{LLM_SCIENTIFIC_DISCOVERY}. The ability to code in different contexts is also fundamental for many of the workflows explored in Section \ref{sec:workflows}.


\subsubsection*{Multi-modality - enriching materials characterization}
\label{sec:multimodal}
The tokenization, positional embedding and attention mechanics of transformers are heavily flexible and therefore capable of jointly modelling different modalities and tasks\cite{MULTIMODAL_SURVEY}.
A prominent example of this is OpenAI's CLIP (Contrastive Language–Image Pre-training)\cite{CLIP}, where a model is trained to maximize the similarity of text and image representations for text-image pairs collected from the internet. The success of CLIP and other multimodal representations\cite{BLIP} has led to the rise of Vision Language Models (VLMs) like GPT-4\cite{GPT4_REPORT}, LLaVa\cite{LLAVA} and Gemini\cite{GOOGLE_GEMINI} which can use information from text and images to aid in the generation and processing of both.

Joint text-image reasoning has the potential to be a useful analysis tool when combined with existing datasets of materials images and descriptions - consider searching the literature for  microstructures that display similar features, defects or artefacts to yours, with potential answers from related papers signposted.

Images are not the only mode of data that transformers can learn to use with text. There are examples using videos via 3D CNN embeddings\cite{VIDEOBERT}, speech/audio using spectrograms\cite{AUDIO_TRANSFORMER} and even graphs via Graph Neural Network (GNN) embeddings\cite{GRAPH_TRANSFORMER_SQL}. Notably, OpenAI's Sora\cite{sora2024} has extended this versatility further by generating high-fidelity videos, demonstrating the application of transformers beyond static images to dynamic, temporal data.

Finding suitable embeddings for the wide range of characterization techniques that exists in materials science (CLIP for micrographs, GNNs for crystallographic information from XRD, 1D CNNs/LTSMs for spectral data) and fine-tuning a transformer or LLM with them could be a promising direction for injecting domain-specific knowledge or priors.

\subsection{Resulting workflows}
\label{sec:workflows}
These properties are flexible and composable, meaning they can be combined in a wide range of potential workflows in various domains, including materials science\cite{HACKATHON_LLM}. Below are a few examples of such workflows, and though they are split into separate sections there are strong links and similarities between them. The key commonality is letting LLMs act as high-level managers whilst other, more robust systems perform low-level tasks.

\subsubsection*{RAG: generation from custom datasets}
\label{sec:RAG}
Retrieval Augmented Generation (RAG) involves performing a lookup into a traditional database and using the retrieved information as part of a prompt to an LLM in order to achieve better or more accurate generation\cite{RAG}.
The lookup is usually based on some function of a user request - one common way to match the semantics of a user's search to a database is `vector search', where embeddings of every item in a database are pre-computed using a language model like BERT\cite{BERT} and the ones with the highest similarity (usually cosine similarity) with the embedding of the user's request are returned\cite{VECTOR_SEARCH_AND_RAG_REEVIEW}.

RAG has several benefits to LLM workflows\cite{RAG_SURVEY}: firstly, hallucinations are reduced as models only need to process existing information in a prompt rather than generate (or fabricate) it. It is more interpretable as the retrieved documents can be linked back to to confirm the results. Finally, these databases can be updated simply by computing the embeddings for the new items - without RAG the LLM would need to be retrained or fine-tuned to add the new information. 

The utility afforded by RAG is clear - many companies are trying to use or sell it \cite{VECTOR_SEARCH_AND_RAG_REEVIEW}, and it is a feature in GPT-4 \cite{GPT4_REPORT}. It is not hard to see how LLMs paired with a vector database of materials papers using, say, MatSciBERT's\cite{MATSCIBERT} text embeddings could prove useful in research. Indeed, some have already used RAG alongside knowledge graphs for materials design\cite{RAG_MATERIALS_DESIGN}.


\subsubsection*{Tool-using and making for analysis pipelines}
\label{sec:tools}

\begin{figure*}
\centering
    \includegraphics[width=0.75\linewidth]{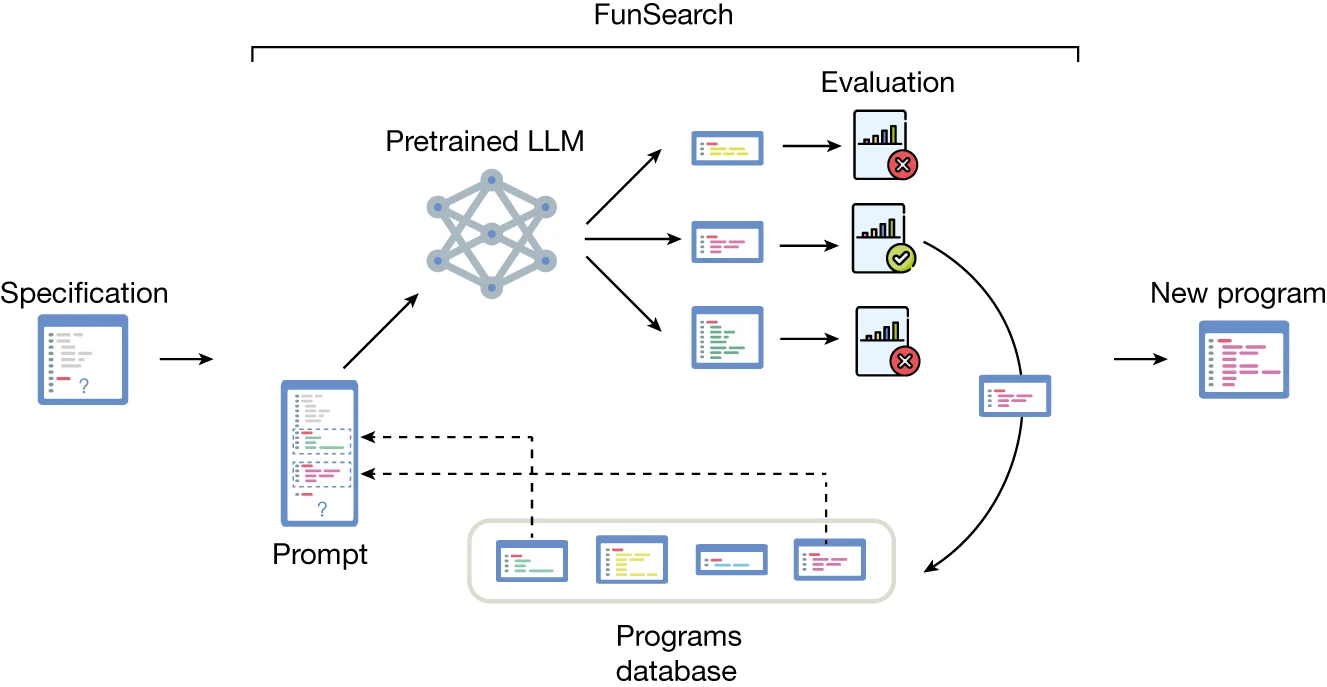}
    \caption{Diagram of the FunSearch\cite{FUNSEARCH} evolutionary workflow, where an LLM is prompted with a problem specification and best example heuristics from the previous iteration and tasked with combining them to generate better candidate heuristics to solve a problem. These new heuristics are evaluated, stored in database and the process repeated. This process was able to discover a new upper bound for the largest cap set in 8 dimensions. Taken from \cite{FUNSEARCH}. }
    \label{fig:funsearch}
\end{figure*}

LLMs can be trained to use use tools like search engines\cite{WEBGPT}, translations, mathematics plugins, \textit{etc.}, which is useful in situations where they typically underperform like arithmetic \cite{TOOLFORMER}. This can be achieved through ICL and `prompt managers' by providing details of the tools and situations in which to use them to the LLM and running the generated code or API (Application-Programmer Interface) calls \cite{VISUAL_GPT, VIPERGPT}. 

Another more involved approach used by Toolformer\cite{TOOLFORMER} was to use ICL to make LLMs annotate an existing language dataset with API calls for a variety of tools where it deemed them useful. They then fine-tuned the model on that data, including a loss term to indicate when the API call improved the accuracy of the generation. This approach has the benefit of not relying on prompts, which can crowd the limited context window and sometimes be ignored by the LLM.

LLMs can generate code and as such are able to produce their own tools. The LLMs As Tool Makers (LATM)\cite{TOOLMAKING} framework used LLMs to generate tools which then be used by other LLMs. They noted that tools were harder to make than use, so had a more powerful LLM (GPT-4) generate the tools, tests and documentation and a weaker LLM (GPT-3.5) use the tools.

A tool-making and using LLM with a human-in-the-loop could be useful for materials science problems where the workflows and requirements are varied (in terms of data types, desired analyses or post-processing) like in image processing. This could be further combined with RAG on relevant papers for domain knowledge engagement and a database of generated tools to obviate the prompt context window limit. Progress has been made on that front, including ChatGPT integration for the ImageJ macro language inside ImageJ itself \cite{CHATGPT_IMAGEJ}.


\subsubsection*{Task integration: the future of automated labwork?}
\label{sec:task_automating}

Various papers have shown that LLMs can act effectively as managers of various sub-components, like other software tools or even other LLMs (called `agents'). This tends to involve feeding the outputs of these tools or LLMs as a prompt into the manager LLM. 

One fun example of co-ordination is `Generative Agents: Interactive Simulacra of Human Behavior'\cite{GENERATIVE_SIMULACRA}, where LLMs acted as villagers in a sandbox with a set of possible actions and locations. They performed inter-agent communication and had a recursively summarised memory of events fed into their prompt to maintain consistency.

Maintaining a memory external to the LLM (\textit{i.e,} in a text file) has been explored by studies like MemGPT\cite{MEMGPT} which aimed to emulate modern Operating System memory management to allow LLMs to perform tasks like large document summarization and multi-session chats. To achieve this they had a traditional scheduler with events for document uploads and timers, and allowed the LLM a set of functions to call in response including send message, read, write, and send interrupts.

`Coscientist'\cite{AUTO_CHEM_RESEARCH} used LLMs as a coordinator to design, plan and execute chemical research. It can call web search APIs, generate and execute Python code, search documentation, interact with and write code for physical hardware. Despite the need for manual intervention to execute the experiment, it is a promising example of how LLMs can orchestrate various research and lab tasks. 

Much effort is being made to integrate LLMs with robotics\cite{ROBOTICS_LLM_SURVERY} as task planners\cite{HOUSEKEEP}, reasoning agents\cite{PALME} or as part of a broader vision-language-action multimodal model\cite{VOLTRON}.
Advancements in grounded robotics and embodied AI will further development of automated labwork, improving all-in-one workflows like Coscientist. However, it is worth noting the margin for error (and hallucinations) is much smaller in labs, where a wide variety of hazardous chemicals and processes are handled frequently.  

\subsubsection*{Optimization loops and `flow engineering'}
\label{sec:flow_engineering}
Another useful LLM-based workflow is meta-optimization, where instead of generating an optimal solution to a problem, an LLM generates the code to produce the optimal solution. `Eureka'\cite{EUREKA} used an LLM to generate reward functions for reinforcement learning applied to robotics simulations. They used a genetic algorithm, where the best generated reward functions and their summary statistics were included in a prompt to allow the LLM to `reflect' and then synthesise a new, better set of reward functions. The framework outperformed expert-written reward functions on a large majority of tasks.

Deepmind's FunSearch\cite{FUNSEARCH} followed a similar approach, using LLMs to generate heuristics for approximating solutions to mathematical problems like the cap set or online bin packing problem. They also used a genetic algorithm framework, asking the LLM to combine aspects of best-performing heuristic programs to generate new ones. Like Eureka, this relied on a combination of ICL, CoT and a feedback signal from an external program - in Eureka's case this was RL simulations using the reward functions which tracked quantities like time upright and for FunSearch this was small validation programs which evaluated how well the heuristic performed (\textit{i.e,} if the cap set was valid and how large it was).

The FunSearch process found a new upper bound for the largest cap set in 8 dimensions, exceeding previous upper bounds found by human mathematicians. Despite this success, this was not a triumph of artificial mathematical understanding - a review of FunSearch noted it was \say{remarkable for the shallowness of the mathematical understanding that it seems to exhibit}\cite{FUNSEARCH_REVIEW} - instead it was proof of the power of LLMs inside an evolutionary framework. 

The LLM in FunSearch did not need to always be correct - the strong feedback signal from the deterministic evaluators ensured mathematical correctness. This is therefore a good model for reconciling the LLM's occasional hallucinations with the need for scientific accuracy.
Based on the results, it seems the key contribution of the LLM was to reduce the search space of the genetic algorithm from all possible functions to all \textit{plausible} functions, hugely increasing convergence time and final performance.

A recent meta-optimization coding paper is AlphaCodium \cite{ALPHACODIUM}, which used a multi-step framework combining reflection on a given specification, human-written tests and LLM-generated tests. The emphasis on tests was because they noticed it was easier to generate useful unit tests (which could then improve future generation) than the correct code. They called this process `flow-engineering' and improved pass accuracy on challenging code problems from 19\% with just GPT-4 to 44\% with GPT-4 as part of the AlphaCodium flow.
A useful feature of all these meta-optimization loops is that they tend to be LLM-agnostic \textit{i.e,} GPT-4 would work equally as well as LLAMA or FALCON.

An important aspect of FunSearch (and other LLM meta-optimizations) was that the programs it generated were interpretable by humans. By examination of the program that generated the new upper bound, the researchers found a new symmetry in the cap set problem. This human-in-the-loop approach to optimization and discovery is appealing in the natural sciences - one could imagine tasking an LLM evolutionary framework to find new functionals in DFT or approximating solutions to physically-relevant combinatorial problems like the max-cut problem\cite{MAXCUT_ISING}. 

\section{LLM workflows in materials science: two case studies}
\label{sec:case_studies}

\subsection{Case study 1: automated 3D microstructure analysis}
\label{sec:case_study_1}

\begin{figure*}
\centering
    \includegraphics[width=0.8\linewidth]{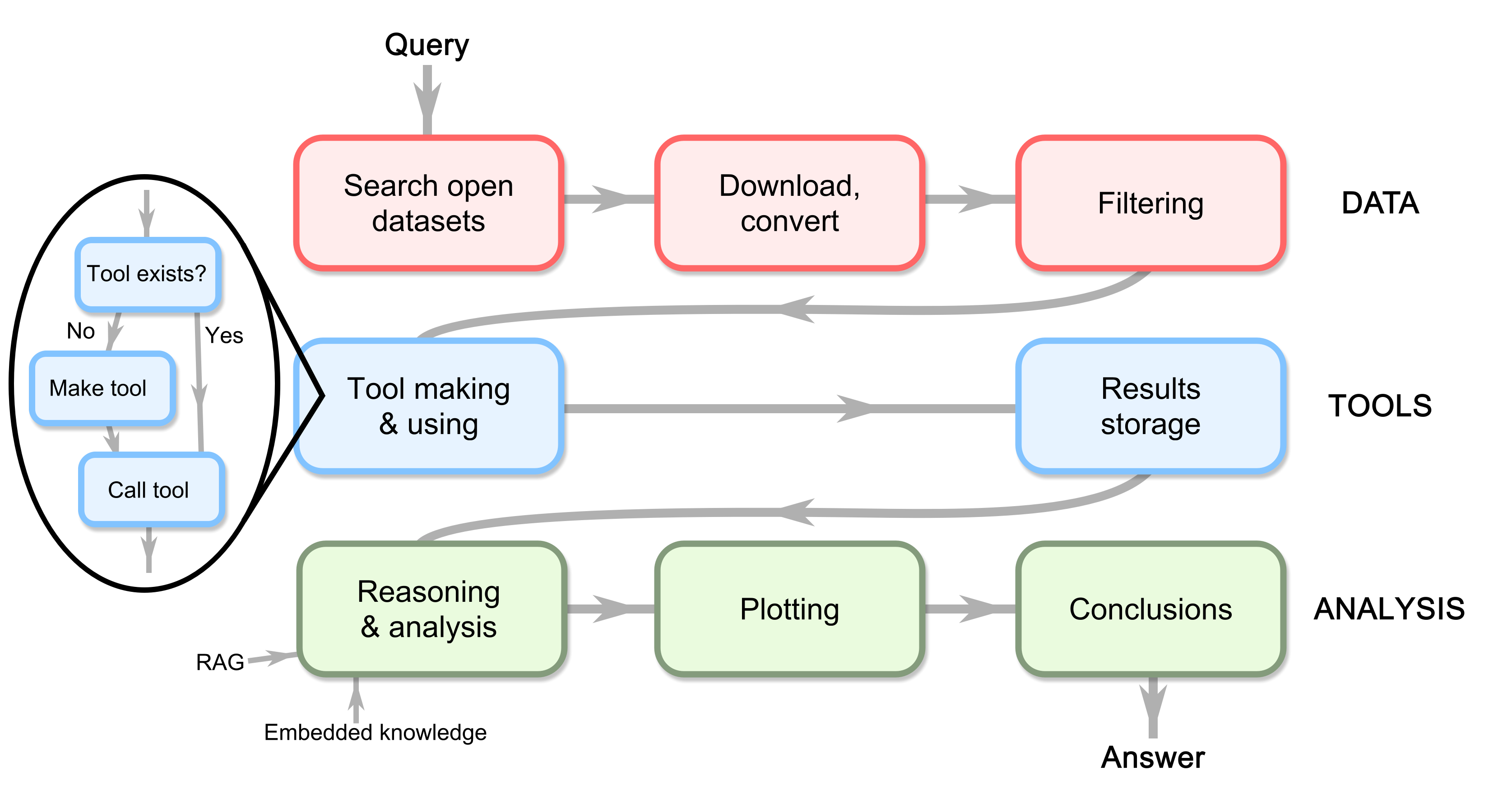}
    \caption{Diagram of MicroGPT's workflow, beginning with dataset collection and filtering. This is followed by tool making and using to extract metrics from the data - this can be an existing tool from its toolkit like tortuosity calculations or created for the specific query. }
    \label{fig:microgpt}
\end{figure*}

In Section \ref{sec:tools} we examined the potential of LLMs to make, use, and orchestrate various tools into automated workflows. Typical materials data analysis pipelines require a combination of domain knowledge, statistical understanding, and various programming skills. The programming required is often non-trivial, involving data handling, conversion, simulations, plotting, \textit{etc.}

LLMs have the potential to reduce the knowledge and skills barriers for these workflows, by offering a natural-language interface to a wide pool of programming knowledge, tool co-ordination, and automation. As an example, we developed \say{MicroGPT} - a specialized chat-bot to streamline 3D microstructure analysis. MicroGPT has a variety of functionalities:

\begin{itemize}
    \item \textbf{Data Acquisition:} MicroGPT can conduct searches for open-source datasets on Zenodo (an interdisciplinary open-access repository) and employ functions to download these datasets using the links available on the respective web pages.
    
    \item \textbf{Filtering:} it can retrieve the dataset's metadata, parse it and subsequently refine the data according to the user's (natural language) specifications. Finally, it organize the filtered data into a newly created file directory. 

    \item \textbf{Integrated simulations:} it can apply simulation tools to the 3D microstructures, documenting the simulation outcomes in formats such as CSV. These results can then be automatically uploaded to a cloud provider given an API key.

    \item \textbf{Data analysis:} it can compare various datasets, collect simulation results and based on user requirements, formulate hypotheses, and provide recommendations.

    \item \textbf{Data visualization: } the results of the data analysis can be plotted, either as histograms for distributions of single properties across the dataset or scatter plots to examine the correlations between properties. 

    \item \textbf{Tool making and reuse:} custom tools can be developed based on the user's specifications, stored and reused in later analyses. Over time this will lead to a library of useful and relevant functions that extend MicroGPT's capabilities. 
\end{itemize}

This was achieved using GPT-4's API. Custom functions were defined in terms of their description, arguments and return values (in \texttt{.json} format) and input to the GPT using OpenAI's `function calling' so the LLM would call them when appropriate. These were implemented in Python and run client-side. We added more system prompts with explicit instructions to improve stability (see Section \ref{sec:prompt_enginering} and Supplementary Section \ref{sec:supp_case_1}).

To demonstrate these functionalities we used MicroGPT to collect and filter data from \say{MicroLib}\cite{MICROLIB}, a collection of plausible, synthetic 3D microstructures generated from DoITPoMS\cite{DOITPOMS} via SliceGAN\cite{SLICEGAN}. It then filtered the structures to only ones related to materials with specific characteristics. Relevant 3D metrics like tortuosity, effective diffusivity, volume fraction, and surface area were calculated using TauFactor 2\cite{TAUFACTOR2} via a function call. 

MicroGPT collated the results, identified a potential outlier, and suggested some materials for further investigation. It successfully correlated metrics such as tortuosity and surface area with desired properties like high flow rates and extensive surface areas for efficient performance.


MicroGPT is a promising example for LLM-assisted analysis workflows, leveraging many of the properties in Section \ref{sec:capabilities} like natural language understanding, programming skills and chain-of-thought reasoning. 
The grounding of MicroGPT using tool like search APIs, RAG, \textit{etc.} is a future research direction which could both reduce factual errors and enhance domain knowledge engagement for reasoning and hypothesis generation. A detailed example dialog and system prompts are available in the Supplementary Section \ref{sec:supp_case_1}.

\subsection{Case study 2: labelled microstructure dataset collection}
\label{sec:case_study_2}
\begin{figure*}
    \noindent\makebox[\textwidth]{
    \includegraphics[width=1.15\linewidth]{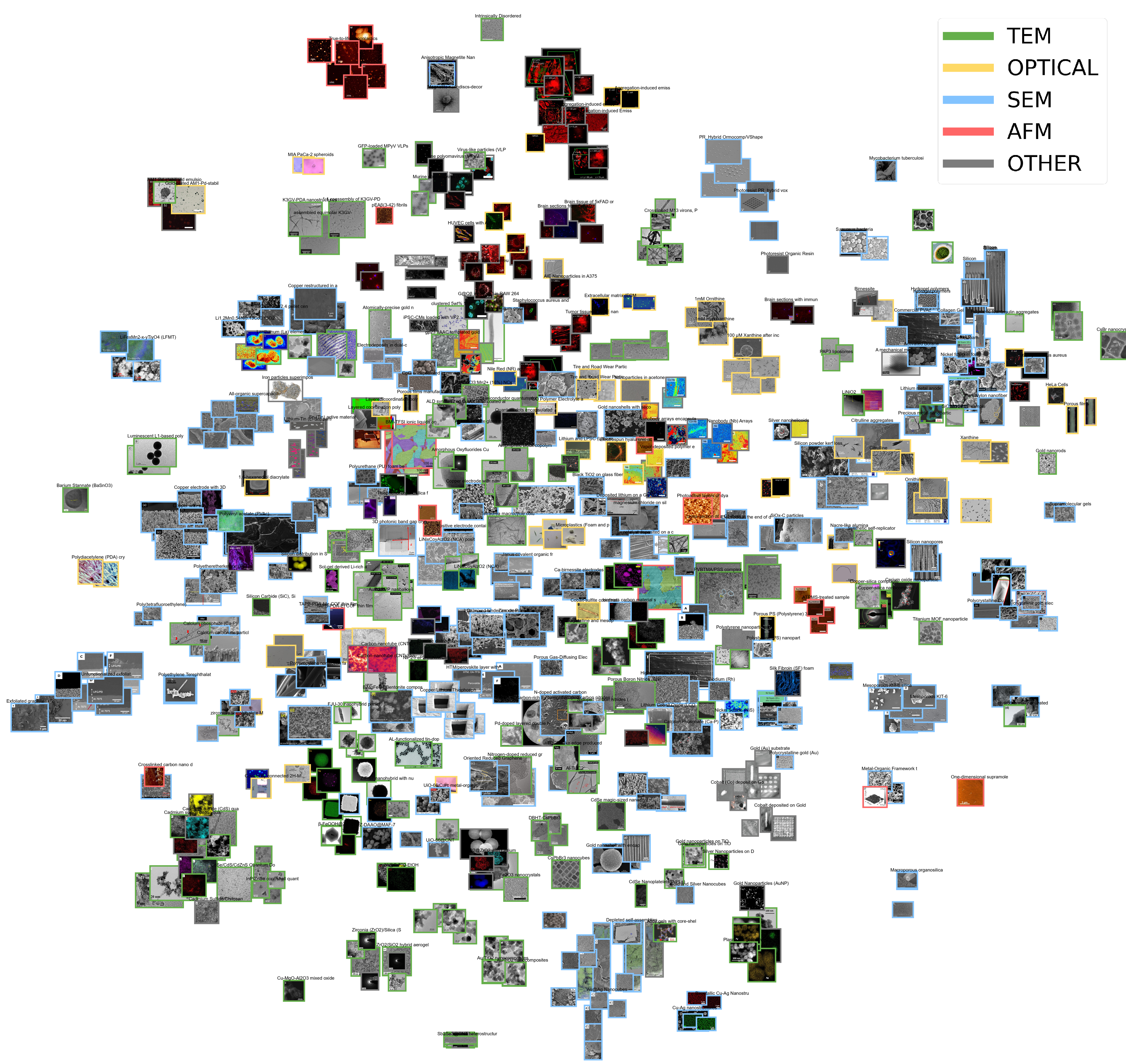}}
    \caption{T-SNE plot of the MatSciBERT\cite{MATSCIBERT} embeddings of the `material' label assigned by the LLM to each micrograph in the dataset based on the paper abstract and figure caption. Border colour denotes the instrument the micrograph was taken with. Similar materials are grouped together: nanoparticles in the bottom right, energy materials in the middle on the left and quantum dots in the bottom-left corner. Best viewed zoomed in.}
    \label{fig:tsne_dataset}
\end{figure*}

There are few large (>1000 entries) micrograph datasets that cover a range of instruments and materials, and even fewer with material-specific labels.
The Cambridge DoITPoMS\cite{DOITPOMS} library contains around 900 labelled micrographs of various materials captured mostly with optical or reflected light microscopy.
Another dataset from Rosella \textit{et. al.}\cite{SEM_22K} contains 22,000 SEM images of materials with taxonomic labels.
Biological datasets are larger and better collated\cite{BIG_BIO_DATA}, contributing to the success of generalist deep-learning approaches like Cellpose \cite{CELLPOSE2}.

Materials science papers contain many high-quality examples of micrographs taken using a variety of techniques, usually with descriptive captions and abstracts.
Traditional string-matching approaches like regex may be capable of detecting whether a given figure contains a micrograph and extracting the instrument used to take it from the caption, but detecting which material is present is generally not possible.
The problem is further complicated if the figure contains multiple sub-figures like plots or diagrams alongside the micrograph, which occurs frequently.

LLMs and VLMs offer solutions to both these problems, displaying strong natural language skills and ability to consider wider contexts like paper abstracts and therefore enabling large-scale automated micrograph collection and labelling from the literature. Some work using GPT-4V for extracting information from a paper's figures exists, for example analysing graphs (PXRD plots, TGA curves, \textit{etc.}) in reticular chemistry papers\cite{IMAGE_MINING_RETIC_CHEM} by treating each page in the \texttt{.pdf} as an image.

We began by scraping paper metadata (title, authors, abstracts, links, \textit{etc.}) from arXiv and chemrXiv that matched the query `microscopy' via their APIs.
For each paper we then downloaded the \texttt{.pdf}, ran the `pdffigures2.0' figure and caption extractor\cite{PDFFIGURES2} and saved the image-caption pairs alongside the metadata.
We further extracted the subfigures for each figure by detecting connected components surrounded by whitespace and removing small (less than 200$^{2}$ px) results.

A two step screening process was used, first we fed captions and abstract to a text-only LLM (GPT3.5 or 4) to determine if a micrograph was present, what instrument was used and what material the micrograph was of.
Next we prompted a VLM (GPT4-V) with the specific subfigure, its parent figure, caption and abstract to work out if that specific subfigure was a micrograph and again what instrument was used and what material was imaged.

\begin{figure}[H] 
    \includegraphics[width=1\linewidth]{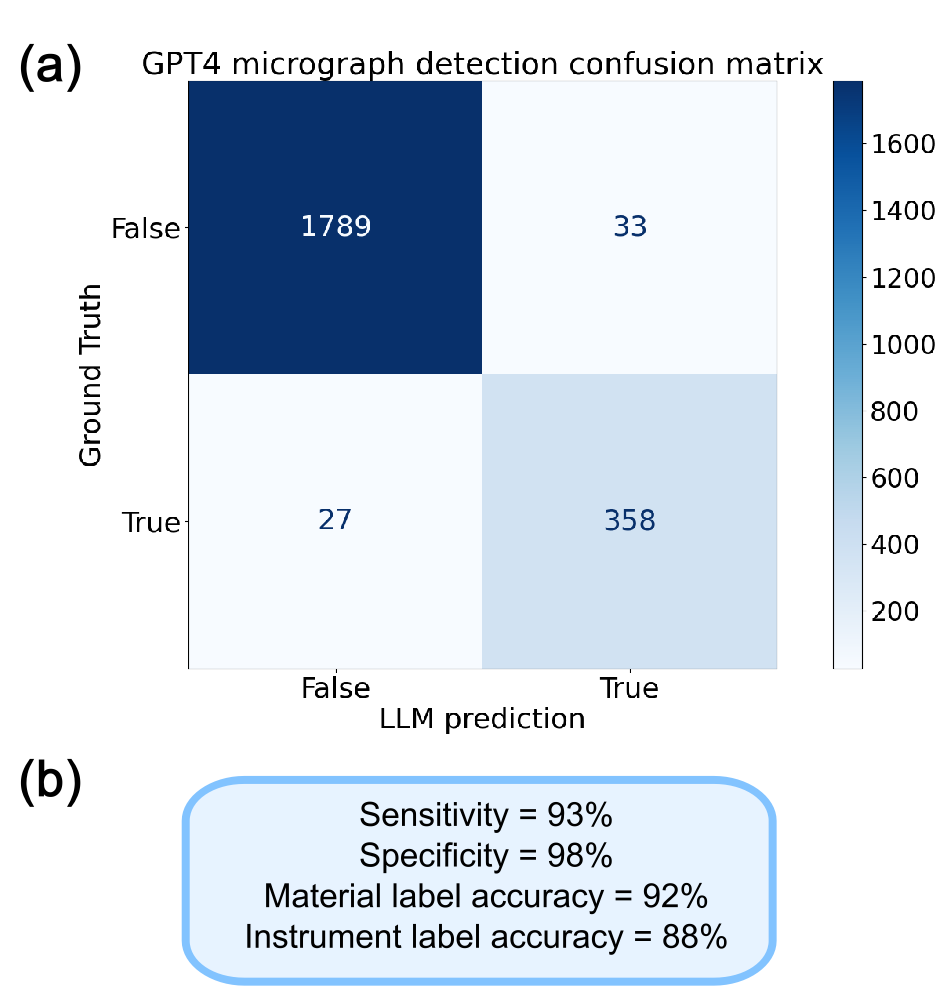}
    \caption{Evaluation of micrograph detection performance of GPT-4 supplied with figure caption and paper abstract, including a confusion matrix in \textbf{(a)} and statistics in \textbf{(b)}. GPT4's performance is strong across the board, with good sensitivity, specificity and accuracy for instrument and material labels.}
    \label{fig:gpt_4_eval}
\end{figure}

After running this process on 382 papers (a subset of the 14,000 scraped) we collected 842 micrographs, each with an instrument and material label - a link to the dataset is available in Section \ref{sec:code}. Figure \ref{fig:tsne_dataset} shows a visualization of the dataset, where micrographs are grouped based on how similar the MatSciBERT\cite{MATSCIBERT} embeddings of their labels are.
The LLM-generated labels were compared to human labels recorded with a custom GUI (developed for this case study) for each figure and subfigure to work out the accuracy of the process.

During the case study we evaluated the performance of various setups, including using GPT3.5 or 4 and whether we prompted the LLM with abstract or not.
GPT4 far outperformed 3.5, and using the abstract led to a minor improvement over not. See Figure \ref{fig:supp_case_study_2_matrices} in Supplementary Section \ref{sec:supp_case_study_2_perf} for details. The performance of GPT-4 with abstract is shown in Figure \ref{fig:gpt_4_eval}, with a sensitivity and specificity above 90\% for micrograph detection, and material and instrument accuracy above 80\%.

We found that LLMs were competent labellers, sometimes matching human labels almost exactly. The success is mostly attributable to the fact that the task could be done with no materials-science specific knowledge due to how well-structured scientific captions are. The text-only LLMs make mistakes when the caption mentions `image' without showing a micrograph, \textit{i.e,} in a plot of statistics taken from an SEM image. The VLM did not have this problem, and there were no false positives after the second step (though this may be because the first step was already a strong filter), this is discussed further in Supplementary Section \ref{sec:supp_case_study_2_eval}.


More details on the setup, including the system prompt, can be found in Section \ref{sec:supp_case_2} of the Supplementary. The code needed to reproduce the results or run on more specific queries is available in Section \ref{sec:code}. In the future we intend to apply this automated approach to a much wider dataset, with the hopes of creating a varied micrograph dataset for computer vision applications.

\section{Issues and challenges}
\label{sec:issues}

There are naturally a few problems with integrating LLMs into materials science workflows, the most prominent and concerning being that of hallucination or confabulation. Huang \textit{et. al.}\cite{HALLUCINATION_SURVERY} provide a taxonomy of hallucination types, separating hallucinations into two main types: factual and `faithfulness'. Factual hallucinations involve being wrong or fabricating facts, and `faithfulness' hallucinations involve ignoring user provided instructions or information or making logical errors. 

Various causes of hallucinations have been suggested\cite{HALLUCINATION_SURVERY}, including (but not limited to) pre-training on incorrect or duplicated data, randomness from output sampling and a `capability misalignment' between the demands made by RLHF fine-tuning and the model's capabilities - LLMs may been trained to hallucinate in some cases.

These fixes for hallucinations exist mostly at the dataset or training level, which is difficult for all but the largest research groups to manage. As noted in Section \ref{sec:RAG}, RAG is a good way to mitigate factual hallucinations\cite{HALLUCINATION_SURVERY}, as manipulating existing data is easier than recall, and it can supply a model with information from outside its training set. Chain-of-thought reasoning can also sometimes mitigate logical hallucinations\cite{HALLUCINATION_SURVERY}, though asking a model to correct itself requires knowing the output was wrong in the first place, reducing the value-add of LLMs.

As well as contributing towards hallucinations, data duplication (alongside autoregression and the pre-training objective) can also contribute to an LLM's tendency to output towards a generic or modal answer. This is not just a problem if asking about uncommon materials or analysis techniques but also if using LLMs to explore a hypothesis space, design principles or automate experiments. The risk of using LLMs in research is that we reinforce existing biases and overlook unconventional approaches not well-represented in the training data.

There are practical issues to implementing LLMs in materials research. The models are expensive to run if using a cloud provider like OpenAI's API, or if run locally require powerful GPUs with at least 8GB of VRAM (which are also expensive).
Quantizing these models (storing their weights with less floating-point precision) can ameliorate this, at the cost of slightly diminished accuracy.
For research groups or companies dealing with sensitive or proprietary data there are privacy issues around uploading data to cloud-based LLMs - running local models is a good workaround but requires more know-how.

\section{Conclusion}
\label{sec:conclusion} 
To conclude, we have explored the basic theory behind LLMs, linking their industrial-scale self-supervised pre-training and reinforcement learning-based fine-tuning to their impressive natural language skills. We then examined existing workflows using LLMs, indicating areas where they have been or could be applied to materials science research. Finally we demonstrated two example workflows using LLMs, one for 3D microstructure data analysis co-ordination and another for the automated collection of LLM-labelled micrographs from the literature.

We believe the versatility and emergent properties of LLMs will make them strong tools in an increasingly automated, connected and data-driven research environment. This is doubly true for materials science which must cover a broad range of length-scales, materials and topics. 

At their current stage of development, LLMs are promising tools for accelerating research and exploration, acting as tireless interdisciplinary workers. They must, however, be used with full understanding of their drawbacks - not as oracles or generators of new, deep insights but in workflows that are robust to and that minimise hallucinations. There is an old saying: \say{fire is a good servant, but a bad master}.

\section*{Code Availability}
\label{sec:code}
The code needed to run the micrograph scraping, extraction and LLM labelling (including the resulting dataset) is available at \url{https://github.com/tldr-group/micrograph_extractor} with an MIT license agreement.

The code to run Microgpt is available at \url{https://github.com/tldr-group/Microgpt}

\section*{Acknowledgements}
This work was supported by funding from Lee Family Scholarships (received by GL), and funding from the the EPRSC and SFI Centre for Doctoral Training in Advanced Characterisation of Materials (EP/S023259/1 received by RD).

The authors would like to thank other members of the TLDR group for discussions and feedback, specifically Isaac Squires who suggested using LLMs to collate a labelled micrograph dataset.

Thank you to arXiv for use of its open access interoperability.


\section*{Competing interests}
The authors declare no competing interests.

\section*{References}
\addcontentsline{toc}{section}{References}

\def\addvspace#1{}

	\renewcommand{\refname}{ \vspace{-\baselineskip}\vspace{-1.1mm} }
	\bibliographystyle{ieeetr} 
    \bibliography{main}

\end{multicols}

\newpage
\section*{Supplementary Information}

\setcounter{section}{0}
\renewcommand*{\theHsection}{S.\the\value{section}}

\makeatletter
\renewcommand \thesection{S\@arabic\c@section}
\renewcommand\thetable{S\@arabic\c@table}
\renewcommand \thefigure{S\@arabic\c@figure}
\makeatother

\newtcolorbox{coloredblock}{
  colback=backcolour,
  colframe=backcolour,
  arc=0pt,
  outer arc=0pt,
  boxsep=10pt,
  fontupper=\ttfamily\footnotesize,
  breakable 
}

\section{Extended LLM theory}

\subsection{Attention}
\label{sec:supp_attn}

Attention is implemented with the matrix multiplication \cite{ATTN_IS_ALL_YOU_NEED}
\begin{equation}
    \text{Attention}(Q, K, V) = \text{softmax}(\frac{QK^{T}}{\sqrt{d_{k}}})V
\end{equation}
where Q, K, V are all linear projections of the input sequence of tokens $X$ with learnable weight matrices $W_{Q}$, $W_{k}$ and $W_{v}$ and $\sqrt{d_{k}}$ is a scaling factor to prevent the dot product getting too large.
Vaswani \textit{et. al.}\cite{ATTN_IS_ALL_YOU_NEED} describe it as \say{a weighted sum of the values, where the weight assigned to each value is computed by a compatibility function of the query with the corresponding key}.
$QK^{T}$ describes how important each token in the context is at predicting or representing the value of every other token\cite{FORMAL_TRANSFORMERS}.
This can be interpreted as a differentiable lookup table\cite{TRANSFORMERS_FROM_SCRATCH} or learned message passing, with $QK^{T}$ determining the similarity of the input to the (learned) keys of this table and V being the (learned) values. The values in $V$ are then weighted by this similarity score and passed to the next layer.

The authors also proposed multi-head attention\cite{ATTN_IS_ALL_YOU_NEED}, where different heads perform the attention calculation on the same input sequence (again in parallel), which improved the model's ability to focus on multiple positions in a sequence. Without the multiple heads, averaging over different training examples where different positions in the sequence contain the relevant pattern (noun-pronoun, verb-subject, \textit{etc.}) would have diminished this ability.

\subsection{Transformer network architectures}
The original transformer network\cite{ATTN_IS_ALL_YOU_NEED} had encoder-decoder architecture, with the encoder formed of $N$ identical encoder layers and the decoder formed of $N$ identical decoder layers. Each encoder layer has multi-headed attention applied to an embedding (vector representation of the tokens in a new subspace) of the input sequence or the previous layer's output, followed by a feed-forward linear layer. The decoder layer has an additional multi-headed attention layer at the start, and `cross-attention' with the output of $n^{th}$ encoder layer. An example encoder layer is show in Figure \ref{fig:theory}.

The original transformer network was an encoder-decoder, which worked well for sequence-to-sequence translation, but not all transformers or LLMs are encoder-decoders. The GPT family \cite{GPT_1, GPT_3, GPT4_REPORT} are decoder-only, which work better for auto-regressive token generation.

\subsection{LLM output sampling}
\label{sec:supp_sampling}

Various sampling and post-processing strategies exist, which can be combined with temperature and each other. `Top-k sampling' chooses the best k-possible tokens and only uses them in the softmax, reducing the computational overhead. `Top-p sampling' chooses from tokens to sample from by summing the $n$ most probable tokens until their cumulative probability is greater than some cutoff $p$ \cite{OUTPUT_SAMPLING}.

`Test-time sampling' generates multiple outputs for the same prompt and uses some heuristic like total sequence probability to decide which to return to the user \cite{OUTPUT_SAMPLING}. This is effective but increases the total compute cost. Other ways of improving results (at the cost of added compute) are search strategies, where various next-token options are explored in parallel. A notable example of this strategy is `beam search'\cite{BEAMSEARCH}.

LLM outputs can be forced to conform to a given format or scheme (like JSON) via `constraint sampling', which involves filtering output token logits to a list of valid tokens for the scheme and selecting the most probable (\textit{i.e,} a JSON should end in a \}) \cite{OUTPUT_SAMPLING}. This form of sampling could prove incredibly useful for writing programs or robotics routines or outputting niche data formats (common in research) that must fit some existing rules. OpenAI have already added a `JSON mode' to their API for such situations.


\section{Case study 1: automated 3D microstructure analysis}
\label{sec:supp_case_1}
\subsection{Full Prompts}
Below are the system prompts for MicroGPT

\vspace{10pt}Prompt 1: MicroGPT system prompt
\begin{coloredblock}
instructions\_microgpt = """ You are an assistant to analyze microstructure. Remember:\\\\
    1. You can invoke tools for analysing tomographic data.
       For image analysing, please ensure to call the function once for each path name provided.\\
       Typically, the number of times the function needs to be invoked corresponds directly to the number of path names you have.\\
    2. After writing the code, always use a function, create\_and\_execute\_python\_file, to upload and execute it.\\
    3. If the user ask for anlysis the all images in a specific folder, please use data\_analysis function. If use ask for analysis an image, please use other function.\\
    4. If the user ask to filter data in a dataset, eg. try to find iron related 3D images in a specific directory, please use data\_filter function.\\
    5. If the user requests to reuse a tool that is included in a Python file, please employ the 'tool\_reuse' function"""
\end{coloredblock}

\vspace{10pt}Prompt 2: simulation function
\begin{coloredblock}
delimiter = "\#\#\#\#"\\\\
system\_message = f"""
    Follow these steps to answer the customer queries.\\
    
    Step 1:\{delimiter\} First, determine if the user is asking a question about analyzing 3D images in a specific directory. If the user is asking about analyzing 3D images in a specific directory, call a function to extract the filenames of images in the directory.

    Step 2:\{delimiter\} Next, call a simulation function to analyze the images.

    Step 3:\{delimiter\} Finally, store all the data in a CSV file.\\

    Use the following format:\\
    Step 1:\{delimiter\} <step 1 reasoning>\\
    Step 2:\{delimiter\} <step 2 reasoning>\\
    Step 3:\{delimiter\} <step 3 reasoning>\\

    Make sure to include \{delimiter\} to separate every step.
    """
\end{coloredblock}

\vspace{10pt}Prompt 3: data filtering function
\begin{coloredblock}
delimiter = "\#\#\#\#"\\\\
    system\_message = f"""
    Follow these steps to answer the customer queries. \\

    Step 1:\{delimiter\} First, confirm whether the user is asking you to filter the data in the database based on their criteria. \
        Determine the directory of the database.\\
    Step 2:\{delimiter\} If the user is asking about \
                filter 3D images in a dataset, you have a function called find\_json, you can use it to unfold zip file, find the meta data of the dataset in the file and extract the metadata\\
    Step 3:\{delimiter\} Now that you have the metadata, which is in a JSON format, focus on the description and keywords within the metadata. Filter all the data in the database that aligns with the user's criteria.\\

    Use the following format:\\
    Step 1:\{delimiter\} <step 1 reasoning>\\
    Step 2:\{delimiter\} <step 2 reasoning>\\
    Step 3:\{delimiter\} <step 3 reasoning>\\

    Make sure to include \{delimiter\} to separate every step.
    """
\end{coloredblock}

\vspace{10pt} Prompt 4: tool reuse function
\begin{coloredblock}

delimiter = "\#\#\#\#"\\\\
    system\_message = f"""
    Follow these steps to answer the customer queries. \\

    Step 1:\{delimiter\} First, confirm whether the user wants to reuse the tool created before, by modifing the code of a file according to his needs and then run it. If yes, call the read\_file function to read this file, specifying the file path clearly \\
    Step 2:\{delimiter\} Now that you have the code, make modifications according to the user's requirements.\\
    Step 3:\{delimiter\} Call the create\_and\_execute\_python\_file function to save and overwrite the original code, keeping the file name unchanged,clearly specifying the file path to be saved.\\
\\
    Use the following format:\\
    Step 1:\{delimiter\} <step 1 reasoning>\\
    Step 2:\{delimiter\} <step 2 reasoning>\\
    Step 3:\{delimiter\} <step 3 reasoning>\\

    Make sure to include \{delimiter\} to separate every step.
    """

\end{coloredblock}

\vspace{10pt}
\subsection{An example interaction}

The following is a conversation between the user and MicroGPT regarding the Microlib Dataset. These dialogues   the following capabilities of MicroGPT: data collection, custom tool creation and reuse, data filtering, data simulation, data analysis, data visualization, etc.\vspace{10pt}

Data Collection

\begin{coloredblock}

User: Can you search for the Microlib online, which is a dataset of 3D microstructures?\vspace{10pt}

Assistant: The Microlib dataset of 3D microstructures is available on Zenodo. Here are some of the relevant records:

1. [Zenodo Record 7118559](https://zenodo.org/records/7118559)\\
2. [Zenodo Record 4047774](https://zenodo.org/records/4047774)\\
3. [Zenodo Record 7867413](https://zenodo.org/records/7867413)\\
4. [Zenodo Record 8112442](https://zenodo.org/records/8112442)\\
5. [Zenodo Record 7660542](https://zenodo.org/records/7660542)\\
6. [Zenodo Record 7043464](https://zenodo.org/records/7043464)\\
7. [Zenodo Record 5011434](https://zenodo.org/records/5011434)\\
8. [Zenodo Record 5090057](https://zenodo.org/records/5090057)\\
9. [Zenodo Record 7702179](https://zenodo.org/records/7702179)\\
10. [Zenodo Record 7820168](https://zenodo.org/records/7820168)\\
You can visit these links to find more information about each dataset and access the data.\vspace{10pt}

User: Download the first one zenodo record 7118559'\vspace{10pt}

Assistant: Do you want to download the file from \\
https://zenodo.org/records/7118559/files/\\
microlibDataset.zip?download=1? (yes/no): yes\\ 
File downloaded: microlibDataset.zip \\
The dataset from Zenodo record 7118559 has been downloaded successfully as `microlibDataset.zip`. If you would like to proceed with extracting and organizing the files, or any other action with this dataset, please let me know!

\end{coloredblock}
\vspace{10pt}
Custom Tool Creation and Reuse

\begin{coloredblock}
User: Please write and execute a script to unzip the file './microlibDataset.zip\vspace{10pt}

Assistant: Python script saved as created\_script.py
Do you want to execute the script? (yes/no):yes\\
I have written and executed a Python script to unzip the file 'microlibDataset.zip'. The contents have been successfully extracted. If you need to work with these files or have any other requests, please let me know!\vspace{10pt}

User: Please unzip the file named './Test\_tool\_reuse.zip' using the tool in './created\_script.py\vspace{10pt}

Assistant: \\
============================Thinking============================\\
Step 1:\#\#\#\# To unzip the file './Test\_tool\_reuse.zip', we first need to read the existing Python script located at './created\_script.py'. We will use the read\_file function to read the content of this file. This will allow us to understand the current functionality of the script and modify it to include the code necessary for unzipping the file. 

Step 2:\#\#\#\# After reading the file, we need to modify the code to add the functionality to unzip files. We will import the 'zipfile' module and use the 'ZipFile' class along with the 'extractall' method to unzip the file. The code modification will include handling the path of the zip file and the destination directory where the contents will be extracted.

Step 3:\#\#\#\# Once the code has been modified to include the unzip functionality, we will call the create\_and\_execute\_python\_file function to save and overwrite the original code in './created\_script.py'. This will update the script with the new functionality, and then we will execute the script to perform the unzipping operation.\\
========================Start solving the problem!=================\\
Python script saved as created\_script.py
Do you want to execute the script? (yes/no): yes\\ The files have been extracted to: ./Test\_tool\_reuse

The script has been updated to unzip the file 'Test\_tool\_reuse.zip', and the execution of the script was successful. The files have been extracted to a directory named 'Test\_tool\_reuse'
\end{coloredblock}

\vspace{10pt}
Data Filter

\begin{coloredblock}
User: In the 'microlibDataset.zip' file, can you find all the 3D images related to cast iron?\vspace{10pt}

Assistant: \\
============================Thinking============================\\
Step 1:\#\#\#\# The user is asking to filter the data in the database based on their criteria, which is to find all the 3D images related to cast iron. The directory of the database is indicated as 'microlibDataset.zip'.

Step 2:\#\#\#\# Since the user is asking to filter 3D images in a dataset, I will use the function find\_json to unfold the 'microlibDataset.zip' file, find the metadata of the dataset in the file, and extract the metadata.

Step 3:\#\#\#\# Now that I have the content of metadata in json-format, I will focus on the description and keywords in the metadata, and filter all the data in the database which align with the user's criteria, which in this case is "cast iron".\\
========================Start solving the problem!=================\\
Found file: data\_anns.json. Is this the metadata file? (yes/no):yes\\
JSON file contains items with the following fields:\\
name\\
link\_doitpoms\\
data\_2D\\
data\_2D\_original\\
data\_3D\\
preview\\
inpaint\_movie\\
long\_movie\\
data\_type\\
brief\_description\\
keyword\\
category\\
element\\
technique\\
long\_description\\
contributor\\
organisation\\
Enter the fields you want to use to filter your data (separated by commas):name,keyword

Based on the metadata extracted from the 'microlibDataset.zip' file, we can filter all the data related to "cast iron." Here are the entries that match the criteria:

- Microstructure 066: Keywords including "cast iron"
- Microstructure 068: Keywords including "cast iron"
- Microstructure 072: Keywords including "cast iron"
- Microstructure 360: Keywords including "cast iron"
- Microstructure 365: Keywords including "white cast iron"
- Microstructure 368: Keywords including "white cast iron"
- Microstructure 370: Keywords including "white cast iron"
- Microstructure 372: Keywords including "white cast iron"
- Microstructure 376: Keywords including "white cast iron"
- Microstructure 378: Keywords including "grey cast iron"
- Microstructure 381: Keywords including "grey cast iron"
- Microstructure 387: Keywords including "grey cast iron"
- Microstructure 393: Keywords including "cast iron"
- Microstructure 396: Keywords including "cast iron"
- Microstructure 406: Keywords including "cast iron"
- Microstructure 782: Keywords including "cast iron"
- Microstructure 784: Keywords including "cast iron"
- Microstructure 797: Keywords including "cast iron"

These entries are filtered out because they contain the keyword "cast iron" or a variation thereof, such as "white cast iron" or "grey cast iron." If you need more information about these microstructures or if you need to extract specific files related to these entries, please let me know.

Please input an exmaple of data filename: microstructure001.tif

Based on the example filename 'microstructure001.tif' provided by the user, we can deduce that the data filenames in the dataset follow the pattern 'microstructureXYZ.tif', where 'XYZ' represents a three-digit number, covering all the ones in the dataset.

Files with the pattern mentioned above have been extracted successfully. However, since I do not have direct access to the file system to list all the filenames, I will now list the filenames based on the metadata provided earlier that are related to "cast iron":

1. microstructure066.tif
2. microstructure068.tif
3. microstructure072.tif
4. microstructure360.tif
5. microstructure365.tif
6. microstructure368.tif
7. microstructure370.tif
8. microstructure372.tif
9. microstructure376.tif
10. microstructure378.tif
11. microstructure381.tif
12. microstructure387.tif
13. microstructure393.tif
14. microstructure396.tif
15. microstructure406.tif
16. microstructure782.tif
17. microstructure784.tif
18. microstructure797.tif

These files are stored in a data folder and are ready for any further analysis or review that you may need. If you'd like to proceed with any specific analysis or have other requests, please let me know!

\end{coloredblock}

\vspace{10pt}Data Simulation

\begin{coloredblock}

User: Could you analyze the 3D images in the './data' folder to determine their tortuosity, diffusion factor, volume fraction, and surface area?\vspace{10pt}

Assistant: 
\\=========================Thinking...=======================\\
Step 1: \#\#\#\# The user is asking to analyse 3D images in a specific directory ('./data'). Therefore, I will call a function to extract the filenames of images in that directory.\\
Step 2: \#\#\#\# Now that we have the filenames, I will call a simulation function to analyse the images for tortuosity, diffusion factor, volume fraction, and surface area.\\
Step 3: \#\#\#\# After the analysis is complete, I will store all the data into a CSV file.
\\=================Start solving the problem!================\\
The image paths from the specified directory ('./data') are as follows:

- './data/microstructure368.tif'
- './data/microstructure378.tif'
- './data/microstructure782.tif'
- './data/microstructure370.tif'
- './data/microstructure365.tif'
- './data/microstructure387.tif'
- './data/microstructure797.tif'
- './data/microstructure360.tif'
- './data/microstructure376.tif'
- './data/microstructure396.tif'
- './data/microstructure372.tif'
- './data/microstructure406.tif'
- './data/microstructure066.tif'
- './data/microstructure072.tif'
- './data/microstructure784.tif'
- './data/microstructure393.tif'
- './data/microstructure068.tif'
- './data/microstructure381.tif'

The simulation results for the analysis of the 3D images are as follows:

1. Microstructure368.tif:
   - Effective Diffusivity: 0.4152
   - Tau (Tortuosity Factor): 1.8422
   - Volume Fraction: [0.2351, 0.7649]
   - Surface Area: 0.0919

2. Microstructure378.tif:
   - Effective Diffusivity: 0.3918
   - Tau: 1.9773
   - Volume Fraction: [0.2252, 0.7748]
   - Surface Area: 0.1074

3. Microstructure782.tif:
   - Effective Diffusivity: 0.3223
   - Tau: 1.7827
   - Volume Fraction: [0.4255, 0.5745]
   - Surface Area: 0.0396

4. Microstructure370.tif:
   - Effective Diffusivity: 0.0464
   - Tau: 4.2417
   - Volume Fraction: [0.8031, 0.1969]
   - Surface Area: 0.0838

... (summary continues for each image)

One of the images, Microstructure797.tif, encountered an issue and could not be processed.

The data from the analysis has been successfully stored into a CSV file. You can download it using the following link:

[Download microstructure\_analysis\_results.csv]\\
(sandbox:/mnt/data/microstructure\_analysis\_results.csv)

\end{coloredblock}

\begin{longtable}{cccccc} %
\caption{Cast iron related microstructure's simulation results in Microlib} \\ %
\toprule
Microstructure & Effective Diffusivity & Tortuosity & Surface Area & \parbox{2cm}{\centering Volume Fraction\\Phase 1} & \parbox{2cm}{\centering Volume Fraction\\Phase 2} \\
\midrule
\endfirsthead

\toprule
Microstructure & Effective Diffusivity & Tortuosity & Surface Area & \parbox{2cm}{\centering Volume Fraction\\Phase 1} & \parbox{2cm}{\centering Volume Fraction\\Phase 2} \\
\midrule
\endhead

\bottomrule
\endfoot

\begin{filecontents*}{tables/data_tau.csv}
Microstructure,Effective Diffusivity,Tau,Surface Area,Volume Fraction Phase 1,Volume Fraction Phase 2
microstructure368.tif,0.4152,1.8422,0.0919,0.2351,0.7649
microstructure378.tif,0.3918,1.9773,0.1074,0.2252,0.7748
microstructure782.tif,0.3223,1.7827,0.0396,0.4255,0.5745
microstructure370.tif,0.0464,4.2417,0.0838,0.8031,0.1969
microstructure365.tif,0.1709,2.0907,0.0429,0.6427,0.3573
microstructure387.tif,0.208,2.8765,0.1266,0.4016,0.5984
microstructure360.tif,0.024,5.9349,0.0657,0.8575,0.1425
microstructure376.tif,0.1422,2.5106,0.1341,0.6431,0.3569
microstructure396.tif,0.6924,1.1616,0.0425,0.1958,0.8042
microstructure372.tif,0.0012,85.388,0.0855,0.9013,0.0987
microstructure406.tif,0.3071,1.7043,0.0908,0.4765,0.5235
microstructure066.tif,0.1644,2.1616,0.0919,0.6447,0.3553
microstructure072.tif,0.2527,2.4606,0.1444,0.3783,0.6217
microstructure784.tif,0.0751,5.1606,0.1768,0.6122,0.3878
microstructure393.tif,0.148,2.7249,0.0614,0.5966,0.4034
microstructure068.tif,0.2682,2.17,0.0898,0.418,0.582
microstructure381.tif,0.4692,1.7356,0.1085,0.1858,0.8142
\end{filecontents*}
\csvreader[
    late after line=\\,
    before reading={\catcode`\#=12},
]{tables/data_tau.csv}{}
{\csvcoli & \csvcolii & \csvcoliii & \csvcoliv & \csvcolv & \csvcolvi}
\end{longtable}

\vspace{10pt}Data Analysis

\begin{coloredblock}
User: Review and compare these data to identify any data points that are conspicuously unreasonable or significantly divergent from the rest.\\

Assistant: Based on the analysis of the dataset for outliers, we've identified the following:

1. **microstructure396.tif**: The effective diffusivity of 0.6924 is identified as an outlier, as it is significantly higher than what we would expect compared to other data points.

2. **microstructure372.tif**: This microstructure has an extraordinarily high tortuosity (Tau) value of 85.388, which is significantly higher than any other data point, indicating that this value is an outlier and warrants further investigation.

3. **microstructure784.tif**: The surface area value of 0.1768 is marked an outlier on the higher end, meaning it is significantly larger than most other surface area values, possibly indicating a highly complex interface between phases.

The rest of the data points appear reasonable, with volume fractions summing close to 1 as they should. It's important to remember that while the statistical method used here highlights significant deviations from the mean, it cannot conclusively determine whether a value is truly unreasonable without additional context regarding the materials and processes involved. These identified outliers should be reviewed carefully with consideration of the experimental conditions or possibly re-measured if experimental error is suspected.\\

User: I believe that only the data for microstructure372 is incorrect. Given that there is an issue with this data, please disregard microstructure 372 when addressing the following question. Analyze which microstructure possesses the highest tortuosity.\\

Assistant: After correctly excluding the data for microstructure372, the microstructure with the highest tortuosity is microstructure360, which has a tortuosity (Tau) value of 5.9349.\\

User: Please compare microstructure 393, 368, and 365\\

Assistant: Here is a comparison of the specified microstructures based on their properties:

\#\#\# Effective Diffusivity (D\_eff):\\
- **microstructure393.tif**: Has a moderate effective diffusivity of around 0.148.\\
- **microstructure368.tif**: Exhibits the highest effective diffusivity among the three, at about 0.415.\\
- **microstructure365.tif**: Close to microstructure393.tif, with a value of around 0.171.\\
\#\#\# Tortuosity (Tau):\\
- **microstructure393.tif**: Has the highest tortuosity with a value of approximately 2.725.\\
- **microstructure368.tif**: Has the lowest tortuosity, with a value close to 1.842.\\
- **microstructure365.tif**: Has a tortuosity value around 2.091.\\
\#\#\# Surface Area:\\
- **microstructure393.tif**: Has a moderate surface area valued at approximately 0.061.\\
- **microstructure368.tif**: Has the highest surface area with a value of around 0.092.\\
- **microstructure365.tif**: Has the lowest surface area, near 0.043.\\
\#\#\# Volume Fraction Phase 1:\\
- **microstructure393.tif**: Has a reasonably high volume fraction phase 1 of approximately 0.597.\\
- **microstructure368.tif**: Has the lowest volume fraction phase 1 of about 0.235.\\
- **microstructure365.tif**: Has the highest volume fraction phase 1, around 0.643.\\
\#\#\# Volume Fraction Phase 2 (complementary to Phase 1):\\
- **microstructure393.tif**: Has the lowest volume fraction phase 2 of about 0.403.\\
- **microstructure368.tif**: Has the highest volume fraction phase 2 of approximately 0.765.\\
- **microstructure365.tif**: Has a moderate volume fraction phase 2, around 0.357.\\
Overall, **microstructure368.tif** stands out due to its high effective diffusivity and surface area, which are beneficial for a catalyst carrier that facilitates reaction and filtration. However, **microstructure365.tif** also has desirable properties, such as the highest volume fraction phase 1, which might imply a more solid structure possibly beneficial for mechanical stability in a catalyst support. **Microstructure393.tif** appears to be a middle ground with moderate properties across the board.
\end{coloredblock}

Data Visualization

\begin{coloredblock}

User: Disregard the unreasonable data, can you generate some figures to create visualizations for the data? Histograms for each numerical column to understand the distribution of values. Scatter plots to explore relationships between pairs of numerical variables (e.g., Effective Diffusivity vs. Tortuosity)\\

Assistant: Here are the visualizations for the data:

\#\#\# Histograms for Each Numerical Column
- The histograms show the distribution of values for each numerical variable: Effective Diffusivity, Tortuosity, Surface Area, Volume Fraction Phase 1, and Volume Fraction Phase 2.

\#\#\# Scatter Plots for Relationships Between Pairs of Numerical Variables
- The scatter plots explore relationships between pairs of numerical variables, such as:

  1. Effective Diffusivity vs. Tortuosity \\
  2. Effective Diffusivity vs. Surface Area\\
  3. Effective Diffusivity vs. Volume Fraction Phase 1\\
  4. Tortuosity vs. Surface Area\\
  5. Tortuosity vs. Volume Fraction Phase 1\\
  6. Surface Area vs. Volume Fraction Phase 1\\
These figures provide insights into how the various microstructural properties are distributed and how they may relate to one another within the dataset.
\end{coloredblock}

\begin{figure}[H]
    \centering
    \includegraphics[width=15cm]{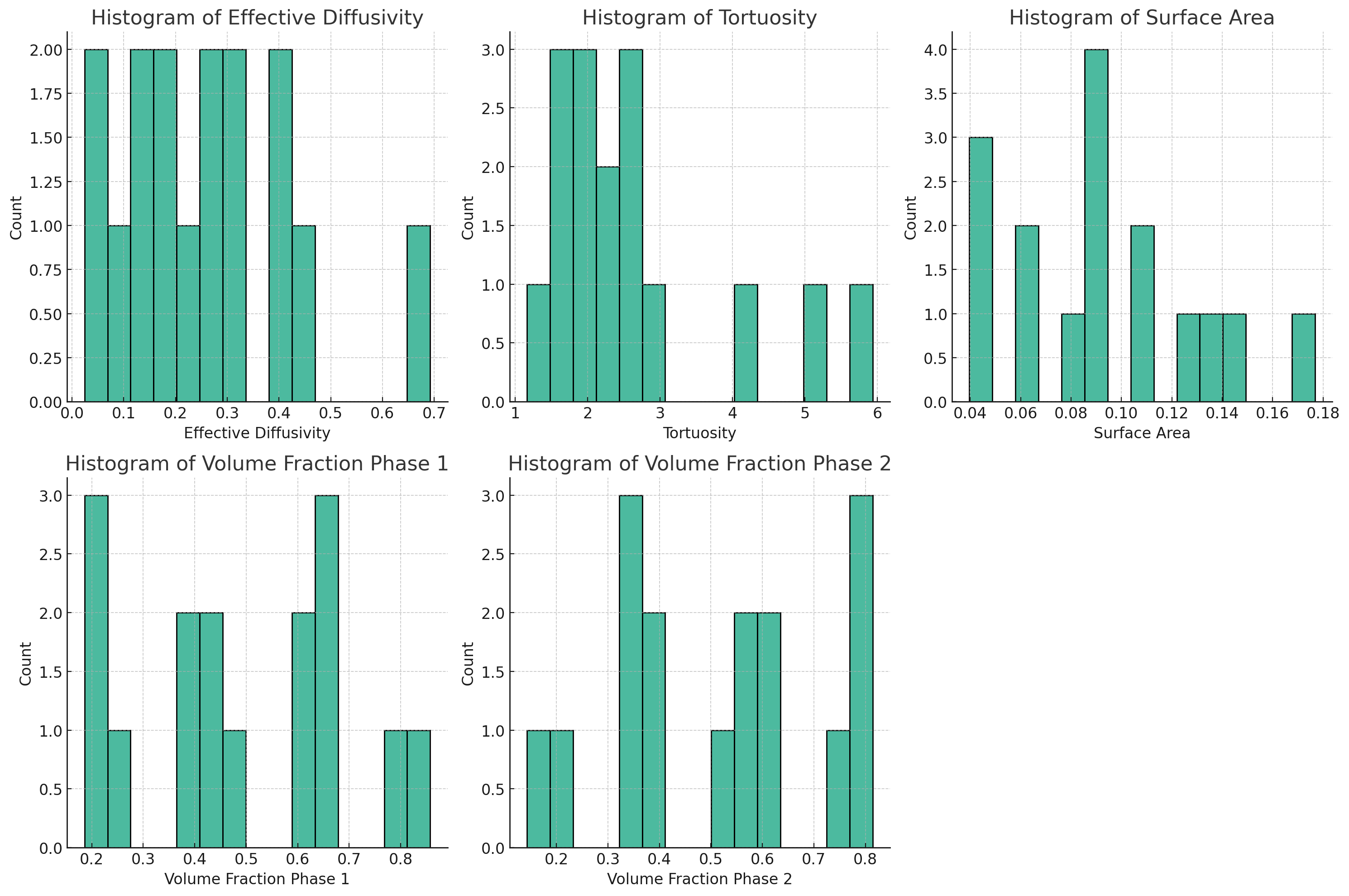}
    \caption{Histograms of effective diffusivity, tortuosity, surface area, and volume fraction of cast-iron microstructures in Microlib as returned by MicroGPT.}
    \label{fig:supp_case_study_1_histogram}
\end{figure}

\begin{figure}[H]
    \centering
    \includegraphics[width=15cm]{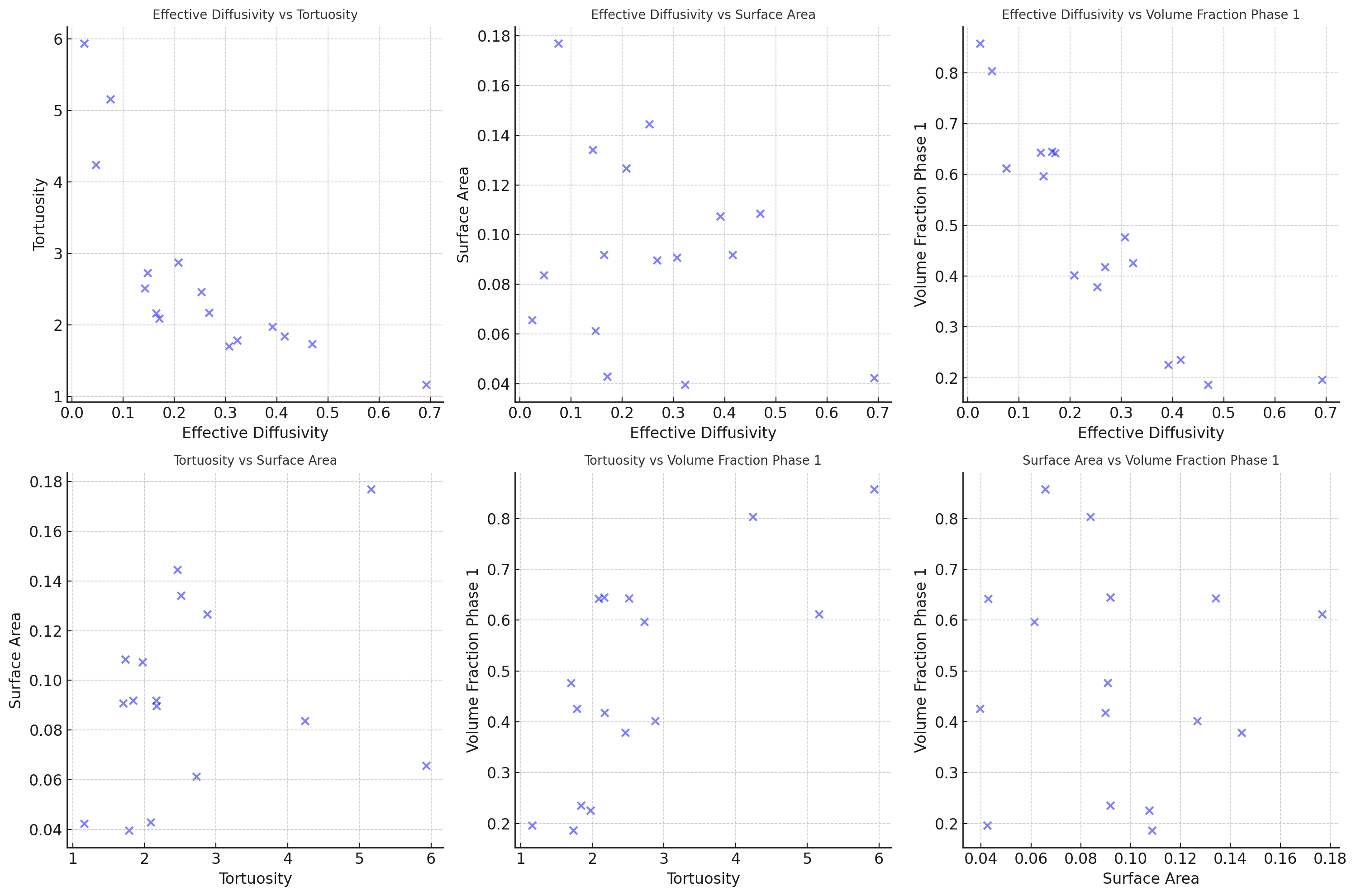}
    \caption{Scatter plots exploring the relationships between pairs of numerical variables of cast-iron microstructures in Microlib as returned by MicroGPT.}
    \label{fig:supp_case_study_1_scatter}
\end{figure}

\section{Case study 2: automated labelled microstructure dataset collection }
\label{sec:supp_case_2}

\subsection{Scraping and figure extraction}
The metadata (title, abstract, authors, DOI, \textit{etc.}) of 14,000 papers matching the search query \texttt{\say{microscopy}} were scraped from arXiv and chemrXiv via their APIs. A subset of 500 of these were downloaded from chemrXiv and the figures and their matching captions extracted with `pdffigures2.0'\cite{PDFFIGURES2}. This worked well but the tool sometimes failed to extract the whole figure or caption.

Figures were split into subfigures by detecting whitespace pixels (pixels whose intensity $\approx 255$) and thresholding. The binary image was dilated to remove small gaps. The connected components of the non-whitspace pixels were computed and bounding boxes generated - these were then saved as subfigures. Small figures like sub-figure letter captions, were discarded. This process works under the assumption that figures use around 5px whitespace to separate subfigures, which tended to be true in most cases.

\subsection{LLM labelling}
We chose to use GPT3.5 and GPT4 as the LLM and VLM in the study, primarily for convenience and performance. Given the relatively uncomplicated nature of the task, a local LLM would be preferable for running this workflow at-scale. The prompts used for each step are in Section \ref{sec:supp_case_study_2_prompts}.

The caption and abstract text-only GPT4 worked well for detecting and categorization of figures (see Section \ref{sec:supp_case_study_2_perf}), but in order to detect if a specific subfigure was a micrograph a vision model was needed. 

\subsection{Evaluation}
\label{sec:supp_case_study_2_eval}
A custom labelling GUI was written to label figures and subfigures in papers in terms of whether they had a micrograph present, what material was shown and what instrument was used to capture the micrograph. It also allowed comparison of the human labels to the LLM labels for evaluation. We chose to do labelling before evaluation (\textit{i.e,} a two-step process) rather than just reviewing the LLM answers in order to avoid bias. 

For the VLM evaluation, we noted the lack of false positives by examining the figures in the produced dataset and resulting T-SNE plot, and not via the custom GUI. This method did not allow us to report the full confusion matrix.  

\subsection{Performance and issues}
\label{sec:supp_case_study_2_perf}
\begin{figure}[H]
    \centering
    \includegraphics[width=1\linewidth]{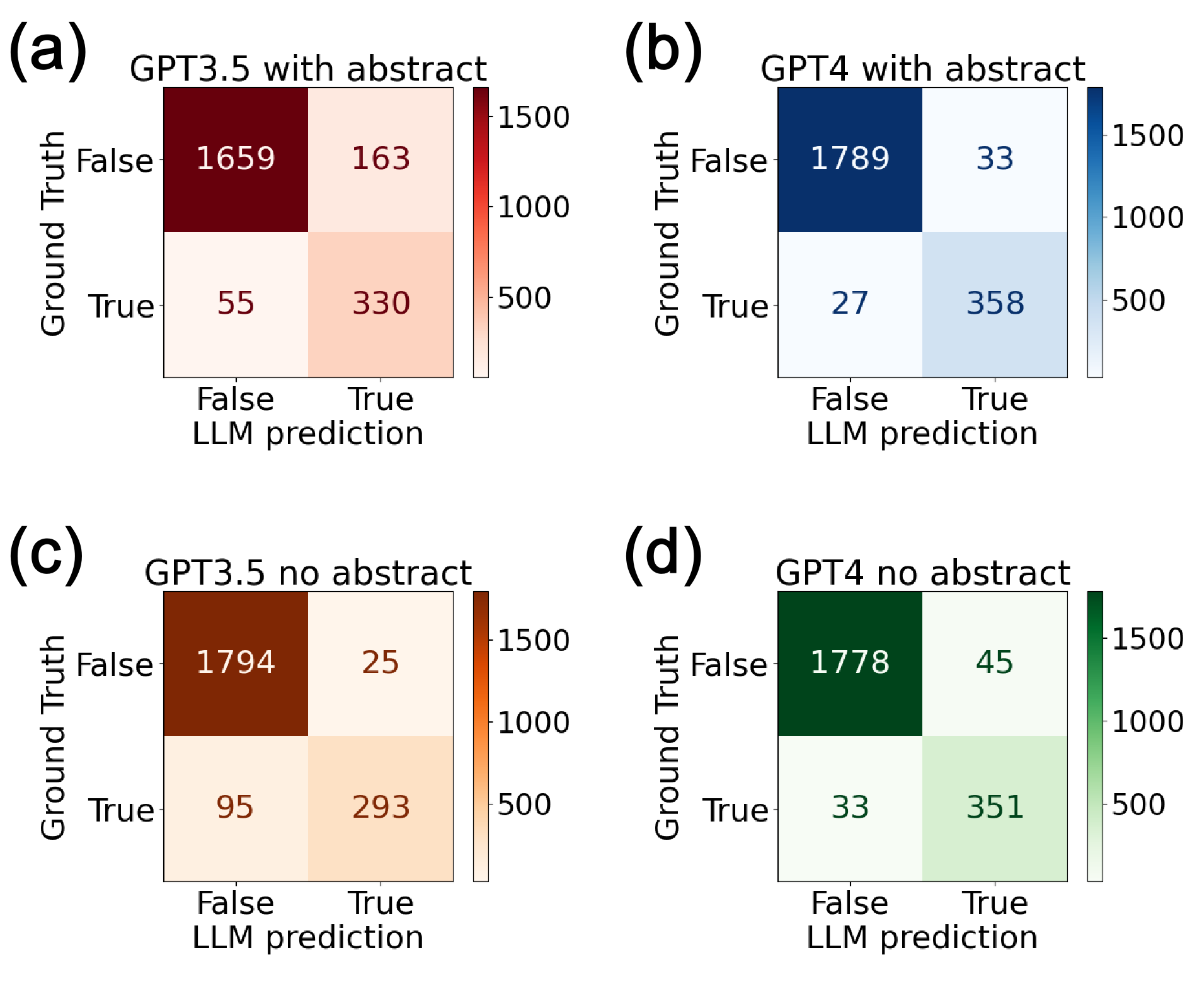}
    \caption{Confusion matrices of micrograph detection for \textbf{(a)} GPT3.5 with caption \& abstract, \textbf{(b)} GPT4 with caption \& abstract,  \textbf{(c)} GPT3.5 with caption \& no abstract and \textbf{(d)} GPT4 with caption \& no abstract. GPT4 with abstract performs the best, with the highest sensitivity and specificity.}
    \label{fig:supp_case_study_2_matrices}
\end{figure}

\begin{figure}[H]
    \centering
    \includegraphics[width=1\linewidth]{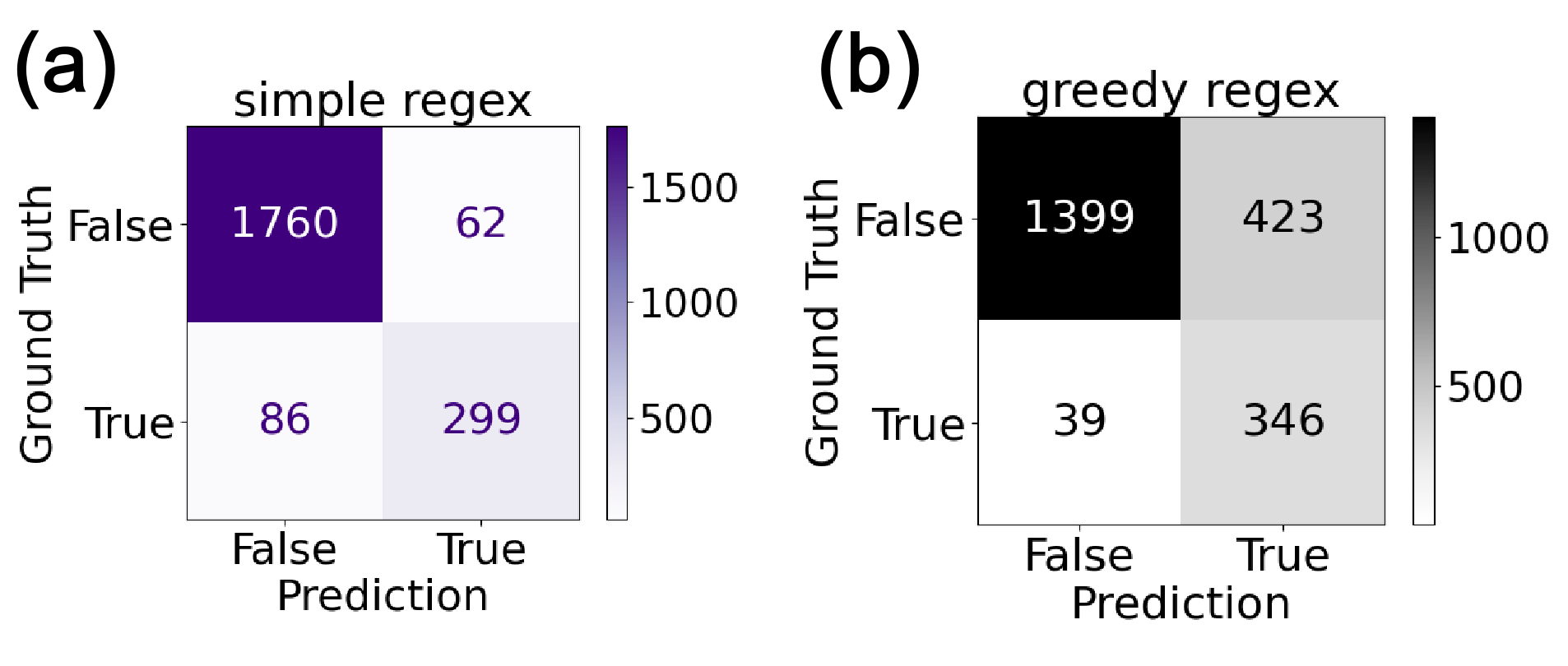}
    \caption{Confusion matrices of micrograph detection for two string matching (regex) approaches. \textbf{(a)} is the simple regex scheme, where a figure is labelled as containing a micrograph if it contains the phrase `image' or `micrograph'. \textbf{(b)} is the greedy scheme, where a figure is labelled as containing a micrograph if it matches the simple scheme or contains the name of a microscopy technique like `SEM' or `AFM'. The simple approach works well though has less true positives than LLM-based results where the caption is more ambiguous. As expected, the greedy approach has more true positives but far more false positives.}
    \label{fig:supp_regex_matrices}
\end{figure}

\subsection{Prompts}
\label{sec:supp_case_study_2_prompts}
\vspace{10pt}Prompt 1: LLM micrograph screening utilizing only the captions of figures as input

\begin{coloredblock}

system\_message ="""\\

    You are an expert on micrographs, images captured using microscopes such as SEM, TEM, or AFM, etc.\\

    Given caption of a figure from an academic paper, answer the question:\\
    Is there a micrograph in this figure? Respond with 'true' or 'false'\\
    
    IMPORTANT: The anwer should only contain pure JSON format only with the fields "isMicrograph"\\

    Here's an example of output:\\

    \{\\
    "isMicrograph": "true or false"\\
    \}\\

    """

\end{coloredblock}

\vspace{10pt}Prompt 2: LLM micrograph screening utilizing the captions of figures and paper abstract as input\\

\begin{coloredblock}

system\_message ="""\\

    You are an expert materials scientist. You study micrographs, which are images taken using a microscope. \\

    Based on an academic paper's abstract and a specific figure's caption, provide answers in JSON format to the following questions:\\
    
    1. Is there a micrograph in this specific figure? Answer with 'true' or 'false'.\\
    2. If a micrograph is present, list the techniques used in this figure (e.g., SEM, TEM, Optical Microscopy). Note that techniques mentioned in the abstract might not be used in this figure.\\
    3. If a micrograph is present, list the the full name of materials depicted in the micrographs e.g., 'Lithium Nickel-Manganese-Cobalt (NMC) 811 cathode' or 'Insulin aggregates'.\\
    4. Are there any noteworthy details about the micrograph, such as unique processing conditions or observed anomalies, in a series of brief phrases (e.g., ['heat-treated', 'cracked', 'sintered'])?\\

    Here's an example of how the JSON output should look with micrograph present:\\

    \{\\
        "isMicrograph": "true",\\
        "instrument": "Technique",\\
        "material": "Description",\\
        "comments": ["comment1", "comment2", "comment3"]\\
    \}\\

    Here's an example of how the JSON output should look without micrograph present:\\

    \{\\
        "isMicrograph": "false"\\
    \}\\
    
    IMPORTANT: The answer should only contain pure JSON data matching the fields provided in the examples.   \\
    """
    
user\_message = f""" The abstract is: \{abstract\_escaped\}, and the captions are: \{captions\_escaped\}"""

\end{coloredblock}

\vspace{10pt}Prompt 3: VLM micrograph recognition with a single sub-figure as input

\begin{coloredblock}

user\_message\_1 = """\\

            You are an expert materials scientist specializing in micrographs. \\
            Typically, these are greyscale and might include annotations like scale bars or arrows, although not always. \\

            Your task:Given a subfigure image from a research paper\\
            1. check if it is a micrograph or another type of figure, such as a graph or diagram\\
            2. check if it includes sub-subfigures.\\
            
            If the image is solely a micrograph without any sub-subfigures, you should respond with 'TRUE'; \\
            otherwise,  respond with 'FALSE'. \\

            Here is an example of output:\\
            "Reason: The image is a micrograph with 2 sub-subfigures.\\
            Answer: FALSE"\\
            
            """

\end{coloredblock}

\vspace{10pt}Prompt 4: VLM micrograph recognition with a main figure and a sub-figure as input

\begin{coloredblock}

system\_message = """\\

            You are an expert materials scientist working on micrographs. The first image is a main image is taken from a research paper. The second image is a subfigure cropped from the main image. It is might be a micrograph. \\

            Focus on the abstract of the paper, captions, and the content of these images. Answer the questions below: \\
            1.Is the cropped image a micrograph? Answer with a single 'true' or 'false'.\\
            2.What technique (e.g., SEM, TEM) was used to create the micrograph in the cropped image? Provide a brief answer, such as 'SEM' or 'Optical Microscopy'.\\
            3.What material is shown in the micrograph? Provide the full name e.g., 'Lithium Nickel-Manganese-Cobalt (NMC) 811 cathode' or 'Insulin aggregates'.\\
            4.If there are any interesting things about the micrograph, like specific processing conditions or anomalies, put these in a list of single phrases (e.g ['heat-treated, 'cracked', 'sintered']). \\
            5.Which part of the caption of the mainfigure does this subfigure correspond to? Extract the caption of the subfigure, and its label e.g (a), (b) if possible.\\

            Here's an example of the JSON output format:\\

            \{\\
            "isMicrograph": True,\\
            "instrument": "Technique",\\
            "material": "Description",\\
            "comments": ["comment1", "comment2", "comment3"]\\
            "subfigure\_caption": "(label): caption of subfigure"\\
            \}\\

            IMPORTANT: The answer should only contain pure JSON data matching the fields provided in the examples.\\
            
            """\\

user\_message = f"The abstract is: \{abstract\_escaped\}, and the captions are: \{captions\_escaped\}"\\

\end{coloredblock}


\end{document}